\documentclass[a4paper,12pt]{article}
\pdfoutput=1 

\usepackage{jheppub} 

\usepackage[T1]{fontenc} 
\usepackage{tensor}
\usepackage{mathrsfs}

\usepackage{mathtools}

\newcommand{\beq}{\begin{equation}}
\newcommand{\eeq}{\end{equation}}
\def\be {\begin{equation}}
\def\ee {\end{equation}}
\def\bs#1\es{\begin{split}#1\end{split}}
\def\ba#1\ea{\begin{align}#1\end{align}}
\def\baed#1\eaed{\begin{aligned}#1\end{aligned}}
\def\bged#1\eged{\begin{gathered}#1\end{gathered}}
\def\bea{\begin{eqnarray}}
\def\eea{\end{eqnarray}}

\def\\nonumber{\nonumber}

\def\G{\Gamma}

\def\pd{\partial}

\newcommand{\cZ}{\mathcal{Z}}

\newcommand{\cK}{\mathcal{K}}

\newcommand{\cR}{\mathcal{R}}

\def\cN{\mathcal{N}}
\def\cV{\mathcal{V}}

\def\Re{\text{Re}}

\def\Tr{\text{Tr}}

\def\pa{\partial}

\def\fr{\frac}

\def\we{\wedge}

\let\foo\bar 
\renewcommand{\bar}[1]{ {\foo{  #1} }{} }

\newlength{\dhatheight}

\title{{\boldmath On  $\alpha'$-effects  from  $D$-branes in $4d \; \cN = 1$}}

\author{Matthias Weissenbacher}

\affiliation{Kavli Institute for the Physics and Mathematics of the Universe, University of Tokyo,\newline Kashiwa-no-ha 5-1-5, 277-8583, Japan\,\footnote{The majority of this work was conducted while being affiliated with the Kavli IPMU at the University of Tokyo.
}}

\emailAdd{ matthias.weissenbacher@ipmu.jp}

\abstract{ In this work we study  type IIB  Calabi-Yau orientifold compactifications in the presence of  space-time filling D7-branes and O7-planes. In particular, we conclude that $\alpha'^2 g_s$-corrections to their DBI actions lead to a modification of the  four-dimensional $\cN=1$ K\"ahler potential and coordinates. We focus on the one-modulus case  of the geometric background i.e.\,$h^{1,1}=1$ where we find that the $\alpha'^2 g_s$-correction is of topological nature. It depends on the first Chern form of the four-cycle of the Calabi-Yau orientifold which is wrapped by the D7-branes and O7-plane. This is in agreement with our previous F-theory analysis and provides  further evidence for a potential breaking of the no-scale structure at order $\alpha'^2 g_s$.  Corrected background solutions for the dilaton, the warp-factor as well as the internal space metric are derived. Additionally,  we briefly discuss $\alpha'$-corrections from other D$p$-branes.}

\begin{document} 
\maketitle
\flushbottom

\section{Introduction}

Since the discovery of D-branes \cite{Polchinski_1995,PolchinskiVol2} they have  played a crucial role in model building as they contribute non-Abelian gauge groups.
In particular, type IIB  flux compactification to four-dimensional space-time generically rely on the presence of space-time filling D3/D7-branes \cite{Giddings_2002,Choi_2004}. In order to cancel the positive charges one is required to introduce O3/O7 orientifold planes in the setup \cite{Brunner_2004,Blumenhagen_2007}. Due to their back-reaction on the geometry  one refers to these backgrounds as Calabi-Yau orientifolds.

In the recent years  F-theory vacua gained a lot of attention \cite{Vafa_1996,weig2018tasi,donagi2008model,Grimm_2011} which in the weak string coupling limit  naturally  incorporate the type IIB D7-branes and O7-planes. 

Two main challenges remain in the landscape of type IIB flux vacua.  Firstly, the problem of K\"ahler moduli stabilization. While the Gukov-Vafa-Witten super-potential \cite{Gukov_2000} in the presence of non-vanishing background fluxes generically allows to stabilize the complex structure moduli \cite{Giddings_2002,Choi_2004}, the leading order terms in $\alpha'$ and $g_s$, do not generate a potential for the geometric K\"ahler deformations which is referred to as the no-scale property. Secondly, the de Sitter up-lift a commonly used phrase to describe the generation of a potential in a low energy theory of string theory which admits a global or local de Sitter minimum, see e.g.\;\cite{Cicoli_2018} for a review.

Historically, the first issue was addressed by incorporating non-perturbative effects such as instantons to generate a potential for the K\"ahler moduli, while the de Sitter uplift mainly relied on  exotic objects such as anti $D3$ branes \cite{Kachru_2003}. 
Recently, the swampland de Sitter conjecture \cite{obied2018sitter} has cast doubt on the consistency of those mechanisms. This however, is an ongoing debate see e.g.\;\cite{Palti:2019pca} for a review.

Instanton effects are exponentially suppressed at large volumes  which make them in general sub-leading to the leading  order $\alpha'$-correction to the scalar potential.
The large volume scenario \cite{Balasubramanian_2005,Conlon_2005} carefully balances an $\alpha'^3 g_s^2$-correction to the K\"ahler potential \cite{Antoniadis_1997,Becker_2002,Bonetti:2016dqh} against instanton effects to the superpotential to achieve moduli stabilization. However, this relies on having small cycles in the Calabi-Yau orientifold while maintaining an overall  large volume.
Lastly,  K\"ahler moduli stabilization may be achieved  by the leading order perturbative corrections \cite{Ciupke:2015msa,Weissenbacher:2019bfb} and may potentially induce a natural de Sitter uplift \cite{Balasubramanian_2004,Westphal_2007,Ciupke:2015msa,Weissenbacher:2019bfb}. Latter  allows for the cycles in the internal space  to be  of comparable size.

This provides a strong phenomenological motivation for the study of $\alpha'$-corrections to the K\"ahler potential and coordinates of four-dimensional $\cN=1$ supergravity theories.  A series of our previous works \cite{Grimm:2013gma,Grimm:2013bha,Weissenbacher:2019mef} addressed the study of the leading order $\alpha'$-corrections in F-theory and thus  weakly coupled  IIB vacua. The foundation of those studies is laid by an extensive analysis of dimensional reduction of higher-derivative terms from eleven to three dimensions \cite{Grimm:2014xva,Grimm:2014efa,Grimm:2017pid}. The correction whose potential existence had been elusive ever since \cite{Grimm:2013gma} is of order $\alpha'^2 g_s$ and thus leading to the well known Euler characteristic correction \cite{Antoniadis_1997,Becker_2002}.\footnote{The dimensional reduction  of the Heterotic string  at $\alpha'^2$-order  including K\"ahler deformations was discussed in \cite{Candelas_2017}.} Recent developments \cite{Weissenbacher:2019mef} suggest the existence of another $\alpha'^2 g_s$-correction to the K\"ahler coordinates proportional to the logarithm of the internal volume which if present breaks the no-scale structure. \vspace{0.2 cm}
\newline
This work  provides  further evidence to the existence of both the correction to the K\"ahler coordinates as well as the K\"ahler potential by the taking the type IIB approach.  We  study  Calabi-Yau orientifold compactifications with space-time filling D7-branes and O7 planes \cite{Jockers:2004yj,Grimm:2004uq}, in particular we focus on the gravitational four-derivative $\alpha'^2 g_s$-sector of the DBI effective actions \cite{Bachas:1999um}, respectively.
 Let us emphasize that our starting point is identical to the one  in \cite{Junghans:2014zla}. However, the  approach discussed in \cite{Junghans:2014zla} lacks certain conceptually required steps to allow conclusions on the K\"ahler metric and thus the K\"ahler potential. Namely, the discussion of the perturbation of the internal metric with respect to K\"ahler deformations\footnote{Note that \cite{Junghans:2014zla} uses the absence of an $\alpha'^2 g_s$-correction to the four-dimensional Einstein-Hilbert term  to conclude on the absence of  an $\alpha'^2 g_s$-correction to the K\"ahler metric. Not only is this implication flawed, but by discussing the E.O.M's  one indeed concludes that an  $\alpha'^2 g_s$-correction to the four-dimensional Einstein-Hilbert is present upon dimensional reduction. }, as well as the discussion of the resulting equations of motions which are modified when  $\alpha'^2 g_s$-corrections to the DBI actions are present. Thus \cite{Junghans:2014zla}  fails to identify any  $\alpha'^2 g_s$-correction in the resulting four-dimensional theory.
 
In this work we mainly study the gravitational $R^2$-terms in the DBI action of D7-branes and O7-planes \cite{Bachas:1999um}.  
But we  also briefly discuss $\alpha'$-corrections from other D$p$-branes, in particular D5-branes and O5-planes and moreover D6-branes and O6-planes in type IIA.  A comprehensive study would as well require the discussion of the $F^2_5R$-sector. However, latter to the best of our knowledge has not been discussed in the literature. 
 \vspace{0.2 cm}\newline
This article is structured as follows. In section \ref{theObjective} we set the stage by introducing  the notion of $\alpha'^2 g_s$-corrections to the four-dimensional K\"ahler potential and coordinates. We continue in section \ref{sec-DBI} by reviewing the starting point of our computation, namely $\alpha'^2 g_s$-effects to D7-branes  and O7-planes effective actions. The dimensional reduction is performed  for a single K\"ahler modulus, i.e\, the overall volume in section \ref{sec:One-mod-red}.\footnote{To perform the computations in this work we employ computational  techniques, in particular we heavily rely on the analytic abstract  tensor algebra  libraries  xAct and xTensor \cite{Mart_n_Garc_a_2002,Mart_n_Garc_a_2003,Mart_n_Garc_a_2008}.} Finally, we conclude in section \ref{FtheoryMatch} on the  corrections to the K\"ahler potential and coordinates by comparison  to the F-theory side. Lastly,   in section \ref{D5branes} we  briefly turn to the discussion of $\alpha'$-corrections to the K\"ahler potential  originating from other D$p$-branes, in particular we initiate the study in type IIA. Higher-derivative terms in the DBI action of D5-branes potentially give rise to a novel $\alpha'^3 g_s$-correction to the K\"ahler potential, as well as  a  potential novel $\alpha'{}^{5/2} g_s$-correction from D6-branes in type IIA.

\section{The objective: 4d  K\"ahler potential and coordinates }\label{theObjective}
The main objective  is  to provide further evidence for the presence of the  conjectured $\alpha'^2 g_s$-correction  to the four-dimensional $\cN=1$ K\"ahler coordinates \cite{Weissenbacher:2019mef}. The latter is proportional to  logarithm of the volume of the Calabi-Yau orientifold
\beq
\;\; \sim \;\; \mathcal{Z}  \; \cdot \; \text{log} \hat\cV \;\; .\label{coor_Kcoord}
\eeq
with $\hat\cV$ the Einstein frame volume i.e.\,the volume of the internal manifold $\cV$ equipped with an additional dilaton dependence as
\beq\label{VolumeEinstein}
\hat \cV = e^{-\tfrac{3\phi}{2}} \cV \;\;  , \;\; \; \text{and} \;\;\;  \hat v^i = e^{-\tfrac{\phi}{2}}  v^i \;\;\;,
\eeq
and where $\cZ$  is a topological correction which will be introduced in detail in section \ref{sec-Def-Z}.
Moreover,  $v^i,\,i=1,\dots,h^{1,1}_+$ are the K\"ahler moduli fields in  dimensionless  units of $ 2 \pi \alpha'$, and  $\phi \equiv \phi(x)$ is the dilaton.\footnote{We denote the external  and internal space coordinates with $x^\mu$ and $y^i$, respectively. Note that this is an abuse of notation of the indices $i,j$, which we also use to denote the abstract index on the K\"ahler moduli space, see e.g.\,eq.\,\eqref{VolumeEinstein}. } 
 We present a systematic study of the one-modulus reduction  of the $R^2$-terms in the DBI and ODBI action of D7-branes and O7's in sections \ref{sec:SUSY-BG} and \ref{sec:One-mod-red}.

To set the stage let us begin by reviewing  the K\"ahler potential and coordinates which are to be corrected at sub-leading order. The volume $\hat\cV$ is dimensionless in units of $(2 \pi \alpha')^3$. The K\"ahler potential and  coordinates  are given by
\beq\label{Kpot}
K = \phi - 2\, \text{log}\Big(\hat\cV + \gamma_1   \cZ_i \hat v^i\Big) \;\; ,
\eeq
and
\beq\label{Kccord}
T_ i =  \rho_i  + i\, \Big( \hat \cK_ i +  \gamma_2  \cZ_i \text{log} \hat\cV \Big)\;\; ,
\eeq
respectively.
The axio-dilaton is
\beq
\tau =  C_0 + i  e^{-\phi}  \;\; ,
\eeq
where $\rho_i$ are the real scalars arising from the reduction of the Ramond-Ramond type IIB four-form fields  $C_4$, $C_0$ the type IIB axion and $\gamma_1,\gamma_2$ are real parameters.  The quantity $ \hat \cK_ i = \hat \cK_ {ijk} \hat v^j \hat v^k$ is defined using the intersection numbers $\hat \cK_ {ijk}$.\footnote{The intersection numbers are given by 
\beq
\hat \cK_ {ijk} = \frac{1}{6!} \int_{Y_3} \omega_i \wedge \omega_j \wedge \omega_k \;\; .
\eeq
where $Y_3$ is the internal Calabi-Yau manifold, and $\omega_i,\, i = 1,\dots,h^{1,1} $ are the harmonic $(1,1)$-forms.
} The sub-leading correction $\cZ_i$ are numbers and thus do not depend on the moduli fields, i.e.\,those are topological quantities of the internal space. Let us stress  that throughout this work we  use dimensionless units 
\beq\label{dimlessVolume}
\cV =\cV /(2 \pi \alpha')^3   \; , \; \; \text{and} \;\;\;   v^i  = v^i / 2\pi\alpha' \;\;,
\eeq
 unless specified else-wise. One thus easily infers that the corrections $\cZ_i$ in eq.'s\,\eqref{Kpot}  and \eqref{Kccord} are of order $\alpha'^2 \,g_s$ compared to the leading order term. Moreover let us note  that the K\"ahler coordinates  eq.\,\eqref{Kccord}  in principle may be corrected by other terms at this order \cite{Weissenbacher:2019mef}  all of which however become constant shifts in the one-modulus case. We thus  omit them for simplicity 
 as we focus on the one-modulus case $h^{1,1} = 1$ in the following. From  \eqref{Kpot}  and \eqref{Kccord} one infers that the K\"ahler potential becomes
\beq\label{Kpot1M}
K = \phi  - 2\, \text{log}\Big( \hat\cV + \gamma_1 \cZ \, \hat\cV^{\tfrac{1}{3}}\Big) \;\; ,
\eeq
and  the K\"ahler coordinates result in
\ba\label{Kccord1M}
T &=  \rho +  i\, \Big( 3 \, \hat\cV^{\tfrac{2}{3}} + \gamma_2  \cZ \, \text{log} \hat\cV  \Big) \;\; , \\[0.2cm]
\tau &=  C_0 +  i \, e^{-\phi}  \;\; .\label{axio-dilaton} 
\ea
The kinetic  terms of the  four-dimensional $\cN=1$ supergravity theory are given by
\beq
S= \frac{1}{2 \kappa_4^2} \int R \ast 1 - 2\, G^{I \bar J} d T_I \wedge \ast d \bar T_ J \;\; \,
\text{with} \;\; 
G^{I \bar J}  = \frac{\pd^2  K(T,\bar T)}{\pd T_I \, \pd \bar T_ J}  \;\; ,
\eeq
where $T_I = (\tau, T_i)$ and $G^{I \bar J} $ is the K\"ahler metric. It is convenient to express the K\"ahler potential \eqref{Kpot1M} in terms of the K\"ahler coordinates using that $\cV \gg 1$,  i.e.\,as an expansion to linear order in $\cZ$.  One finds that \eqref{Kpot1M} and \eqref{Kccord1M} become
\beq\label{Kpot-TTbar}
K =- \log\big(-\tfrac{i}{2}\big( \tau - \bar \tau \big)\big) - \Big(3 - \frac{9 i \, \gamma_2 \, \cZ}{T -\bar T} \Big) \cdot \log\big(-\tfrac{i}{6}\big( T - \bar T \big)\big) - \frac{12 i \, \gamma_1 \, \cZ}{T -\bar T} \;\; .
\eeq
Thus in the one-modulus case one infers from \eqref{Kpot1M}, \eqref{Kccord1M}, \eqref{axio-dilaton} and  \eqref{Kpot-TTbar} that
\ba\label{ActionFromKpot}
S = \frac{1}{2 \kappa_4^2} \int R \ast 1 \;\;
 -& \frac{1}{2}  d \phi \wedge \ast d \phi  \; - \;  \frac{1}{2} \, e^{2 \phi}  d C_0 \wedge \ast d C_0 \\ \nonumber
- & \left( \frac{2}{3 \,\hat\cV^{2}  } - \frac{ 8 \gamma_1 + 3\gamma_2 }{9\,\hat\cV^{8/3}}  \cZ   \right) d \hat\cV \wedge \ast d \hat\cV \;\; \\[0.2 cm] \nonumber
-   &  \left( \frac{1}{6 \,  \hat\cV^{ 4/3}} -\frac{ 8 \gamma_1 + 9\gamma_2 }{36\,\hat\cV^{2}}  \cZ  \right)d \rho \wedge \ast d \rho  \;\; .
\ea
Where we have again used the fact that $\cZ \ll \cV$ for geometries where the internal volume is large in units of $(2 \pi \alpha')^3$. The primary goal is to argue that the obtained correction $\cZ$ is indeed of topological nature and moreover agrees with our previous F-theory analysis. For a definition of $\cZ$ see eq.\,\eqref{Z-def}.  Secondly, from \eqref{ActionFromKpot} one infers that by fixing the kinetic terms obtained by dimensional reduction on the internal Calabi-Yau orientifolds in the presence of D7-branes and O7-planes one can fix $\gamma_1$ and $\gamma_2$.
  In particular, we will derive the correction for the case of a Calabi-Yau orientifold with  a single K\"ahler modulus and with eight coinciding D7's and one O7$^-$. Note that this setup is different compared to the one studied in \cite{Weissenbacher:2019mef} where only a single D7-brane is present with the class of the  divisor wrapped being a multiple of the one wrapped by the O7-plane, such that tadpole cancellation is guaranteed.
Lastly, note that  in eq.\eqref{ActionFromKpot}  the  correction to the mixed kinetic terms of the dilaton and the Einstein frame volume as well as the  correction to the kinetic term of the dilaton is absent. 
\newline
{\bf Scalar Potential.} Let us emphasize that the Ansatz for the K\"ahler potential and K\"ahler coordinates breaks the no-scale structure and thus generates a scalar potential for the K\"ahler moduli fields for non-vanishing vacuum expectation value of the super-potential  $W$, given by e.g.\,the flux-superpotential  after stabilizing the complex structure moduli denoted by $W_0$. The F-term scalar potential is given by
\beq
V_F = e^K \big(G_{I \bar J} D^I W \bar D^J \bar W - 3|W|^2 \big) \;\;,
\eeq
where $ D^I W = \partial_{T_I} W + W \partial_{T_I} K $. By using  \eqref{Kpot1M} and \eqref{Kccord1M}  one infers in the one-modulus case that
\beq
V_F = \frac{3 \, \gamma_2 \, e^\phi \, \cZ }{2 \, \hat\cV^{8/3}}  \, |W_0|^2\;\; .
\eeq
\newline
{\bf Comments.}  Let us close this section with some concluding remarks on the Ansatz for the K\"ahler potential and coordinates  \eqref{Kpot1M} and \eqref{Kccord1M}. In particular let us  emphasize that it is the complete Ansatz at  order $\alpha'^2 g_s$  consistent with the functional form of K\"ahler metric which will be derived later in this work. Firstly, note that the real part of the K\"ahler coordinates \eqref{Kccord1M} are protected by shift symmetry of  $\rho$ against $\alpha'$-corrections.\footnote{Note that one may want to write the  general Ansatz for an $\alpha'^2$-correction up to constant shifts as
\beq
\Re T = \rho \cdot \Big( 1 + d_1 \, \cZ \, \frac{1}{\hat\cV^{2/3}} \Big) +  d_2  \, \cZ \, \text{log} \hat\cV  \;\;,
\eeq
with $d_1,d_2$ parameters. However, the first correction  breaks shift-symmetry and the term proportional to the logarithm does not constitute a sub-leading term in the limit $\hat\cV \to \infty$. Thus one concludes that the real part of $T$ is not to be corrected.
} Analogous conclusions hold for the generic modulus case  \eqref{Kpot} and \eqref{Kccord}. Moreover, the imaginary part of the axio-dilaton  eq.\,\eqref{axio-dilaton}  is not expected to be corrected by $\alpha'$-corrections.  We proceed in this work without a general proof of this assumption. However, note that in \cite{Bonetti:2016dqh} we  have explicitly confirmed that in the case of the $\alpha'^3 g_s^2$-correction to the K\"ahler-potential there is no correction to the axio-dilaton.

Lastly, let us emphasize that the possibility of  a correction  to the K\"ahler coordinates  of a $4d, \,\cN=1$ theory proportional to the logarithm of the internal volume has already been discussed in the literature \cite{Conlon_2009,Conlon_2010}.  To establish a correction of the latter to our topological coefficient in eq.\,\eqref{coor_Kcoord} is of great interest.\footnote{Note that \cite{Conlon_2010}  concludes that the correction is of order  $g_s$  compared to the leading term.} Furthermore, it is worth noting that an analog correction  to the K\"ahler coordinates appears in $3d,\,\cN=2$ originating from higher-derivative terms to eleven-dimensional supergravity i.e.\,the low energy limit of M-theory compactified on a Calabi-Yau fourfold \cite{Grimm:2017pid}.  Moreover, there is no symmetry forbidding such a correction \eqref{coor_Kcoord} and thus one would generically expect  it to be present, see e.g.\,\cite{Palti:2020qlc}. Concludingly,  the presence of a correction to the  K\"ahler coordinates as in  eq.'s\,\eqref{coor_Kcoord}, \eqref{Kccord} and \eqref{Kccord1M} is thus likely. We support this conclusion by providing more  explicit evidence in this work.

\section{$\alpha'^2 g_s$-effects to D7-branes  and O7-planes}\label{sec-DBI}
In this section we discuss the relevant D7-brane and O7-plane actions.  Those are are in general composed of the Dirac-Born-Infeld (DBI) as well as the Wess-Zumino\footnote{ Also referred to as Chern-Simons action.} action as
\beq\label{SD7}
 S_{DBI} + S_{WZ}  \;\; .
\eeq
Let us first review the classical  contribution to \eqref{SD7}. One finds the DBI action \cite{Polchinski_1995,Witten_1996,PolchinskiVol2} for a  D7-brane to be
\beq
S^{\tiny (0)}_{DBI} = - \mu_{7} \int  d^7Y \, e^{- \Phi}  \Tr \sqrt{- \text{det}\big( i^\ast (g+ B)_{ij}  + 2 \pi \alpha'F_{ij}\big)} 
\eeq
with brane charge $\mu_7 = ( (2 \pi)^3 \cdot (2 \pi \alpha')^{4})^{-1} $,  $i$ the embedding map of the D7 into the ten-dimensional space-time and $i^\ast$ its pullback. Moreover, $F$ is the gauge field strength on the world-volume of the brane \cite{denef2008les}, $B$ the NS-NS two-form field and $\Phi$ the dilaton.
The leading order Wess-Zumino contribution is given by
\beq
S^{\tiny (0)}_{WZ} = \mu_{7} \int  \Tr \, \big( i^{\ast}\big( C \wedge e^{B}\big) \wedge e^{ 2 \pi \alpha'F} \big)\;\;,
\eeq
where $C = \sum_n C_n$ the sum over the various Ramond-Ramond fields.
The $O7^-$-plane contributions are 
\beq
S^{\tiny (0)}_{O-DBI} =  8 \, \mu_{7} \int d^7Y  \, e^{- \Phi}  \Tr \sqrt{- \text{det}\big( i^\ast g_{MN} \big)} \;\; \text{and} \;\;
S^{\tiny (0)}_{O-WZ} = - 8\,  \mu_{7} \int  i^{\ast} C_8  \;\;.
\eeq
In the following we refer to the $O7^-$-plane simply as $O7$.
At the relevant $\alpha'^2g_s$-order  there are $R^2$-terms in the DBI action of the D7 brane and O7 planes, which we review in section \ref{sec:DBI-R2}.

\subsection{$R^2$-terms in the DBI effective actions of D7's and O7's}\label{sec:DBI-R2}

In this section we discuss the gravitational leading order corrections to the DBI action of D-branes and O-planes.
We focus on the case  of D7-branes and O7-planes.
The Riemann squared DBI action for D7-branes in the string frame  \cite{Bachas:1999um,Fotopoulos_2001,Wyllard_2001,Fotopoulos_2002}  
is given by
\ba\label{DBI}
S_{DBI}^{R^2} =  \tfrac{(2 \pi \alpha')^{2}}{ 192  } \mu_{7} \int_{D7} e^{-\Phi} \Big[& R_T{}_{\alpha \beta \gamma \delta} R_T{}^{\alpha \beta \gamma \delta} - 2 {\bf R_T}{}_{\alpha \beta } {\bf R_T}{}^{\alpha \beta }  \\[0.2cm]\nonumber &- R_N{}_{\alpha \beta a b} R_N{}^{\alpha \beta  a b} + 2  {\bf \bar R}_{ab} {\bf \bar R}^{ab}  \Big] \ast_8 1 \;\; ,
\ea
with brane charge $\mu_7 = ( (2 \pi)^3 \cdot (2 \pi \alpha')^{4})^{-1} $ and
with $ R_T{}_{\alpha \beta } = R_T{}_{\alpha  \gamma  \beta}{}^{\gamma}$.  $R_T$ denotes the Riemann tensor in the tangent directions of the D7-brane and $R_N$  is the normal curvature of the D7-brane. 
The bold notation in \eqref{DBI} refers to the dilaton dependence
 \ba\label{dilaton-DBI}
 {\bf R_T{}_{\alpha \beta }} = R_T{}_{\alpha \beta } + \hat\nabla_\alpha \hat\nabla_\beta \Phi  \;\;\ ,\;\;\;\;\;
 {\bf \bar R_{ab} }=  \bar R_{ab} + \hat\nabla_a \hat\nabla_b \Phi \;\; ,
 \ea
 found in \cite{Jalali:2015xca}, where $\Phi = \Phi (x,y)$ is the ten-dimensional dilaton.
The  object $ \bar R_{ab}$ seems  not to do admit a natural geometric  interpretation on the  D7-brane  however is defined in terms of the total Ricci tensor $\hat R_{ M N}$ of the ten-dimensional space-time and the  second fundamental form $\Omega$ as
 \ba \bar R_{ab} &=  \;\;  g^{\alpha \beta} \hat R^{}_{a \alpha b \beta} + g^{\alpha \beta} g^{\gamma \delta } \delta_{ac} \delta_{bd} \,\Omega^{c} {}_{\alpha \gamma}  \Omega^{d} {}_{\beta \delta} \;\; , \\
 R_T{}_{\alpha \beta \gamma \delta} &=  \;\; \hat R{}_{\alpha \beta \gamma \delta} + \delta_{ab}\big(\Omega^{a} {}_{\alpha \gamma}  \Omega^{b} {}_{\beta \delta} - \Omega^{a} {}_{\alpha \delta}  \Omega^{b} {}_{\beta \gamma} \big) \;\; ,\\
 R_N{}_{\alpha \beta}{}^{  a b} &= \;\; - \hat R { }^{\, a b} {}_{\alpha \beta} + g^{\gamma \delta} \big(\Omega^{a} {}_{\alpha \gamma}  \Omega^{b} {}_{\beta \delta} -\Omega^{b} {}_{\alpha \gamma}  \Omega^{a} {}_{\beta \delta}\big) \;\; ,  \label{DBI-eqs}
  \ea
  where $g_{\alpha \beta}$ denotes the metric on the tangent space of the D7 and the $\hat R$ refers to total space Riemann tensor where the respective tangent and normal indices  are pulled back form the total space.
 For  precise  definitions we refer the reader to appendix  \ref{sec:Immersions}.  For  geodesic immersions i.e. $\Omega = 0$ of  D7-branes and O7-planes it was confirmed  \cite{Schnitzer:2002rt} that 
\beq  \label{ODBI}
S_{ODBI}^{R^2} =  2^{p-5} \,  S_{DBI}^{R^2} \;\; ,
\eeq
for a $D_p$-brane on the orientifolded background. It is expected  that this  relation \eqref{ODBI} holds for generic immersions i.e. $\Omega \neq 0$.
\vspace{0.2cm}
\newline
{\bf $F_5^2 R$-sector.}
Lastly, let us comment on the  four-derivative terms which are quadratic in the Ramond-Ramond four-form field $C_4$, with field strength $F_5 = d C_4$.  Relevant for our discussion are terms of schematic form $F_5^2 R$.\footnote{The $(\nabla F_5)^2 $-sector does not contribute in the one-modulus case. } Those can in principle be fixed by six-point open string disk  and projective plane amplitudes, see e.g.\,\cite{Stieberger:2009hq,Mafra:2011nv}. We are not a aware of a derivation of the terms in the DBI or ODBI effective actions.
 Their absence will lead to a free parameter in the reduction result  in section \ref{sec:One-mod-red} which however may be fixed via a match with the F-theory approach as discussed in section \ref{FtheoryMatch}.

\subsection{$R^2$-terms to  the Wess-Zumino  effective actions}\label{sec:WZ_R2}
One finds the  $R^2$-contribution to the Wess-Zumino action for a D7 brane \cite{bachas1998lectures,Cheung_1998,Robbins_2014} to be
\beq\label{R2-WZ_action}
S_{WZ - D7} =  \frac{(2 \pi)^4 \alpha'^2}{48} \mu_{7} \int_{D7} C_4 \wedge \Big( p_1(ND7) - p_1(TD7))\Big) \;\; ,
\eeq
where $p_1(ND7),\,p_1(TD7)$ are the first Pontryagin class  of the  tangent and normal bundle, respectively, and $C_4$ is the Ramond-Ramond four form field strength.  We now use the fact that we consider space-time filling D7-branes, i.e.\,wrapping a four-cycle in the internal space. The latter is  a complex manifold and one thus may relate the previous expression to the first and second Chern-classes as
\beq
p_1 =   c_1^2 - 2 c_2 \;\;.
\eeq
The rank of the normal space is of complex dimension one and thus $c_2(ND7) = 0$. 
The discussion for the O7-planes proceeds analogously \cite{Morales_1999,Stefa_ski_1999}. One finds
\beq\label{action-WZ-O}
S_{WZ - O7}  =  - 2^{7-4} \cdot \frac{1}{4} \cdot S_{WZ - D7}  =  - 2  \cdot S_{WZ - D7}\;\; ,
\eeq
where we used the fact that we are dealing with D7-branes and O7-planes to fix the pre-factor.

\subsection{Embedding of branes in Calabi-Yau orientifolds}\label{sec-Def-Z}

The Calabi-Yau orientifold  is defined as $oY_3 = Y_3 /  \sigma$, where  $\sigma: Y_3 \to Y_3$ in type IIB string theory is an isometric holomorphic involution \cite{acharya2002orientifolds,Brunner_2004,Sen_1996}, i.e. \,$\sigma^2 $ equals to the identity map. The involution  preserves the complex structure and metric from which one infers the action on the K\"ahler form to be
\beq
\sigma^\ast J = J \;\;,
\eeq
where $\sigma^\ast$ denotes the pullback map. For $O7$-planes one infers that $\sigma^\ast \Omega = - \Omega$, with $ \Omega \in H^{(3,0)}$ being the unique holomorphic $(3,0)$-form. 
For the D7-branes to preserve four-dimensional $\cN=1$ supersymmetry the hyper-surface wrapped by it inside $oY_3$ is to be minimal, i.e.\,the representative $D_m$ inside the Homology class which minimizes the volume \cite{Becker_1995}.
The latter can be shown to be equivalent to the  divisor $i: D_m  \xhookrightarrow{}  oY_3$ being a K\"ahler sub-manifold which morover implies that $\Omega^a{}_{\alpha}{}^\alpha = 0$.  The scalar curvature  of $D_m$  generically is non vanishing 
\beq
 R_T |_{oY_3} \neq 0 \;\; ,
\eeq
and by using eq.'s\,\eqref{complexKMetric}-\eqref{DefFirstChern}  may be written  as
\beq\label{R-to-Z}
R_T |_{oY_3} =  4 \pi \cdot \ast{_4} \big(  c_1(D_m) \wedge  \tilde J \big) \;\; ,
\eeq
where $c_1(D_m)$ is the first Chern-form of the divisor and $\tilde J $ its  K\"ahler form  which is inherited from the total space  $ \tilde J = i^\ast J $. 
To avoid introducing yet another notation we  simply denote with $|_{oY_3}$ the restriction of object entirely to the internal  Calabi-Yau space, see appendix \ref{minimal_Immersions} for details.

In the one-modulus case we can  factorize  the volume modulus dependence as $\tilde J =   \cV^{1/3}\omega $. Such that $\omega $  does not depend on the K\"ahler modulus.
For later use let us define the topological quantity
\beq \label{Z-def}
\mathcal{Z} := \int_{D_m} c_1(D_m) \wedge  \omega\;\; .
\eeq

\section{$\alpha'^2 g_s$-corrected Calabi-Yau orientifold  background solution  }
\label{sec:SUSY-BG}
In this section we analyize the E.O.M's resulting from the leading order ten-dimensional type IIB supergravity and the $\alpha'^2 g_s$-corrected DBI actions of D7-branes and O7-planes.
The relevant part of the ten-dimensional type IIB leading order supergravity action in the string frame takes the form\footnote{Where we use the notation $ |H_3|^2 =\tfrac{1}{p!} H_{MNO}H^{MNO}$. }
\beq\label{classicaltypeIIB}
S^0_{IIB - S} = \frac{1}{2\kappa_{10}}\int e^{- 2 \Phi}\Big( \, R \ast_{10} 1 + 4 \, d \Phi \wedge \ast d \Phi  -\frac{1}{2} |H_3|^2\Big) -\frac{1}{2} |F_3|^2-\frac{1}{4} |F_5|^2 \;.
\eeq
where 
\beq
H_3 = dB_2\;\;  , \;\; F_3 = dC_2-C_0 d B_2 \;\; \text{and} \;\; F_5 = dC_4 - \tfrac{1}{2} C_2\wedge d B_2 + \tfrac{1}{2} B_2 \wedge d C_2 \;\;,
\eeq 
are the usual NSNS and RR three-form field strengths and with
\beq
F_5 = \ast_{10} F_5 \;\;,
\eeq
being self-dual. A Weyl rescaling by
\beq
g^S_ {MN} = e^{\Phi /2}g^E_{MN} \;\;,
\eeq
leads to the ten-dimesnioal Einstein frame action
\footnote{
The  Einstein frame action results in
\beq\label{IIBE}
S^0_{IIB - E} \supset \frac{1}{2\kappa_{10}} \int R \ast_{10} 1 \dots\;\; .
\eeq
where we have used \eqref{10dWeyl} }.

Moreover, we have used $g^S,\, g^E$ to denote the string and Einstein-frame metric, respectively and $2 \kappa_{10} = (2 \pi)^7 \alpha'^4 = \mu_7^{-1}$.
The  basic  framework of our discussion are supersymmetric flux compactifications of type IIB super-string theory on a Calabi-Yau threefold \cite{Giddings_2002,Choi_2004,Blumenhagen_2007,Jockers:2004yj,Grimm:2004uq}. 
Setting aside higher-derivative corrections to the ten-dimensional supergravity action  \cite{Giddings:2001yu} the background metric is given by
\ba\label{Ansatz-BG0}
\Phi &= \phi^{\tiny (0)}  \;\;\;, \;\;\;\; \phi^{\tiny (0)}  = \text{const.}  \;\; ,\\[0.2 cm] 
ds^2_S &= e^{\,  2A  }  \eta_{\mu \nu} dx ^\mu dx^\nu + e^{ -2A }   g_{ij}  dy^i d y^j \;\; \nonumber
\ea
 where  $g_{ij}$ denotes the unmodified Calabi-Yau threefold metric. The warp factor $A = A(y)$ determines the background $F_5$-flux via
 \beq\label{WarpandF5}
F_5 = \big(1+\ast_{10}\big) de ^{4A} \wedge dx^0  \wedge dx^1 \wedge dx^2 \wedge dx^4 \;\; .   
\eeq
Moreover, the  flux obeys the Bianchi identity 
\beq\label{BianchiF5}
dF_5 = H_3 \wedge F_3 + \rho_6 \;\; ,
\eeq
where  $\rho_6$ encodes the D3-brane charge density associated with potential localized sources. Integrating \eqref{BianchiF5} one yields the D3-brane tadpole cancellation condition
\beq\label{d3tadpole}
\frac{1}{(2 \pi \alpha ')^2}\int_{Y_3} H_3 \wedge F_3  + N_{D_3}  \;\; = \;\; 0 \;\; ,
\eeq
where $ N_{D_3}$ is the total number of D3 branes. In the presence of D7-branes and O7-planes  the D3 tadpole cancellation condition \eqref{d3tadpole} gets modified \cite{Plauschinn_2009,Blumenhagen_2009} as
\beq\label{d3tadpoleMod}
\frac{1}{(2 \pi \alpha ')^2}\int_{Y_3} H_3 \wedge F_3  + N_{D_3}  \;\; = \;\; \frac{N_{O_3}}{4}  + \sum_{i}^{\# D7's} N^i_{D7} \frac{\chi(D^{\tiny D7}_i)}{24} + \sum_{i}^{\# O7's}  \frac{\chi(D^{\tiny O7}_i)}{12}  
\eeq
\newline
where the sum runs over the  stacks of  D7-branes containing the number of $N_{D7}$  each. The  $D_i$'s are the four-cycles in the Calabi-Yau orientifold $oY_3$ wrapped by the D7's and O7's, respectively. Moreover $\chi$ is the Euler-characteristic of the 4-cycles  and $N_{O_3}$ the number of O3-planes. We have set the background gauge flux of the D7-branes to zero throughout this work.
We do not discuss localized D3 branes and O3 planes in this work.  From \eqref{d3tadpoleMod} one infers that non-vanishing $H_3$ and $F_3$ flux are consistent with the absence of D3 branes as long as the Euler-characteristic of the divisor wrapped by the D7-branes and O7 planes do not vanish.
Moreover, let us note that the D3 and O3 higher-derivative corrections won't affect our results as the latter cannot  affect the K\"ahler metric of the moduli space with the same functional dependence as D7-branes. However, the presence of D3-branes gives rise to other effects \cite{DeWolfe:2002nn}.
Lastly, the tadpole cancellation condition of D7-branes \cite{Plauschinn_2009} is given by 
\beq\label{d7tadpoleMod}
\sum_{i}^{\# D7's} N^i_{D7} \cdot \Big(\, [D^{\tiny D7}_i] + [D^{\tiny D7}{}'_i] \,\Big) + 8 \sum_{i}^{\# O7's}  [D^{\tiny O7}_i]  \;\; = \;\; 0 \;\; ,
\eeq
where $[\cdot]$ denotes the class of the four-cycle wrapped by the D7's and O7's and the prime denotes the orientifold image i.e.\,the action on the background geometry in the presence of the O7-plane.
To accommodate for \eqref{d7tadpoleMod} we choose a setup of eight D7-branes and one O7-plane which wrap the same four-cycle inside the Calabi-Yau fourfold and vanishing gauge flux. Moreover, the Calabi-Yau orientifold admits only one K\"ahler modulus $h^{1,1} = 1$, i.e.\,the overall volume and moreover  one finds that $h^{1,1}_+ = 1$. Although the latter requirement  may seem restrictive it is sufficient to deduce the K\"ahler-potential and  coordinates as those  may be generalized to the generic K\"ahler moduli case, see e.g.\,the original  derivation of the Euler characteristic correction to the $\cN=1$  K\"ahler potential \cite{Becker_2002}. The generic moduli case derivation was done rather recently by \cite{Bonetti:2016dqh}.

To  combine  \eqref{DBI} ,\eqref{ODBI}  and  \eqref{classicaltypeIIB}  one equips the DBI and ODBI action with $\delta_{D7}$, which  is non-vanishing on the world-volume of the D7-branes. The definition of the scalar quantity $ \delta_{D7} $ is simply given by\footnote{Note that \eqref{defDeltaD7} may be alternatively expressed via a 2-form $\hat\omega$  which is Poincare dual to the cycle wrapped by the D7-brane. In other words one finds that $\int_{oY_3} J \wedge  J \wedge \hat \omega = \int_{D7} \ast_8 1$ and thus $ \delta_{D7} \propto   \hat\omega_{ij} g^{\tiny (0)}{}^{ij}$. }
\beq\label{defDeltaD7}
\int \delta_{D7} \ast_{10} 1 = \int_{D_7} \ast_8 1 \;,\;\; \text{and in particular} \;\; \int_{oY_3} \delta_{D7} \ast_{6} 1 = \int_{D_m} \ast_4 1 \;\;.
\eeq
\newline
{\bf Dilaton E.O.M.}
We proceed by deriving the string frame equation of motion for the dilaton. One infers from \eqref{classicaltypeIIB}  and eight-times the DBI action \eqref{DBI} plus the ODBI action \eqref{ODBI}  for eight coincident D7's  on top of a single O7 to be
\ba\label{EOMdilaton}
R^{\tiny(1)} + 4 \nabla^{\tiny(0)} _i  \nabla^{\tiny(0)}{} ^i  \Phi^{\tiny (1)}  -\tfrac{1}{2}| H_3|^2 &+ 2 \alpha\,\delta_{D7}\, e^{\Phi }\cdot  \nabla^{\tiny(0)} {}^{\alpha} \nabla^{\tiny(0)}{}^{\beta} R_T{}_{\alpha \beta}|_{oY_3}  \\ \nonumber
 &- 2  \alpha \,\delta_{D7}\, e^{\Phi }\cdot \nabla^{\tiny(0)}{}^{a}\nabla^{\tiny(0)}{}^{b} \bar R_{ab} |_{oY_3} + \dots = 0 \;\; ,
\ea
where we have used the fact that the leading order solution eq.\,\eqref{Ansatz-BG0} denoted by the superscript $ ^{\tiny(0)}$  is Minkowski times internal Calabi-Yau, and moreover that the classical dilaton solution is constant. For notational simplicity  we have defined 
\beq
\alpha = 12 \cdot \frac{(2 \pi \alpha ')^2}{192}  \;\; .
\eeq
The ellipsis in \eqref{EOMdilaton} denote  $\mathcal{O}(\alpha^2)$ contributions as well as terms quadratic in the Riemann-tensor of the internal space. 
\newline
{\bf External Einstein equation.}
Let us next turn to derive the Einstein equations from \eqref{classicaltypeIIB} , \eqref{DBI}  and \eqref{ODBI}.\footnote{Note that there is no contribution from the classical DBI and WZ actions due to tadpole cancellation.}  The external Einstein equation  is given by
\ba\label{EinsteinExternal}
 R^{\tiny(1)} + 4 \nabla^{\tiny(0)} _i  \nabla^{\tiny(0)}{} ^i  \Phi^{\tiny (1)}  -\tfrac{1}{2}| H_3|^2 &-\tfrac{1}{2} e^{2\Phi}| F_3|^2  \\[0.2cm] \nonumber
 +\, 4 \alpha\,\delta_{D7}\,  e^{\Phi }\cdot  \nabla^{\tiny(0)} {}^{\alpha} \nabla^{\tiny(0)}{}^{\beta} R_T^{\tiny(0)}{}_{\alpha \beta}|_{oY_3}& - 4  \alpha \,\delta_{D7}\, e^{\Phi }\cdot \nabla^{\tiny(0)}{}^{a}\nabla^{\tiny(0)}{}^{b} \bar R^{\tiny(0)}_{ab} |_{oY_3} + \dots = 0 \;\; .
\ea
\newline
{\bf Internal Einstein equations.}
The internal Einstein equations  split in tangent and normal directions. The internal normal components  are given by
\ba\label{EinsteinNormal}
\delta_{ab}\Big( &\tfrac{1}{2}R^{\tiny(1)} + 2 \nabla^{\tiny(0)} _i  \nabla^{\tiny(0)}{} ^i  \Phi^{\tiny (1)}  -\tfrac{1}{4}| H_3|^2-\tfrac{1}{4} e^{2\Phi}| F_3|^2 \Big)  \\[0.2 cm] \nonumber
&- R^{\tiny(1)} _{ab} + \tfrac{3}{2}| H_3|^2 _{ab}+\tfrac{3}{2} e^{2\Phi}| F_3|^2_{ab} - 2  \alpha \,\delta_{D7}\, e^{\Phi }\cdot \nabla^{\tiny(0)}_{\alpha}\nabla^{\tiny(0)}{}^{\alpha} \bar R^{\tiny(0)}_{ab} |_{oY_3} + \dots = 0 \;\; ,
\ea
The tangent components are given by \footnote{Where we have used the notation
\beq
| H_3|^2 = \frac{1}{3!} H_3{}_{ijk} H_3{}^{ijk} \;\; ,\;\;\; | H_3|^2 _{ab} = \frac{1}{3!} H_3{}_{a ij} H_3{}_{b}{}^{ij} \;\; ,\;\;\; | H_3|^2_{\alpha \beta}  = \frac{1}{3!} H_3{}_{\alpha ij} H_3{}_{\beta}{}^{ij}  \;\; ,
\eeq
and analogously for $| F_3|^2 _{ab} $ and $| F_3|^2 _{\alpha \beta}  $ .
}
\ba\label{EinsteinTangent}
&g^{\tiny(0)}_{\alpha \beta}\Big( \tfrac{1}{2}R^{\tiny(1)} + 2 \nabla^{\tiny(0)} _i  \nabla^{\tiny(0)}{} ^i  \Phi^{\tiny (1)}  -\tfrac{1}{4}| H_3|^2-\tfrac{1}{4} e^{2\Phi}| F_3|^2 + 2 \alpha\,\delta_{D7}\,e^{\Phi }\cdot  \nabla^{\tiny(0)} {}^{\gamma} \nabla^{\tiny(0)}{}^{\delta} R_T^{\tiny(0)}{}_{\gamma\delta}|_{oY_3}  \nonumber \\[0.2 cm]  \nonumber
 &- 2  \alpha \,\delta_{D7}\, e^{\Phi }\cdot \nabla^{\tiny(0)}{}^{a}\nabla^{\tiny(0)}{}^{b} \bar R^{\tiny(0)}_{ab} |_{oY_3} \Big) \\[0.2 cm]  \nonumber
&  +\alpha\,\delta_{D7}\, e^{\Phi } \Big(- 4  \nabla^{\tiny(0)}_{\gamma}\nabla^{\tiny(0)}_{(\alpha}  R_T{}^{\tiny(0)}_{\beta) \gamma} |_{oY_3}- 4  \nabla^{\tiny(0)}{}^{\gamma}\nabla^{\tiny(0)}{}^{\delta}  R_T{}^{\tiny(0)}_{\alpha \gamma \beta \delta } |_{oY_3}  + 2 \nabla^{\tiny(0)}_{\gamma}\nabla^{\tiny(0)}{}^{\gamma}  R_T{}^{\tiny(0)}_{\alpha \beta} |_{oY_3} \Big) \\[0.2 cm]  
&- R^{\tiny(1)} _{\alpha \beta} + \tfrac{3}{2}| H_3|^2 _{\alpha \beta}+\tfrac{3}{2} e^{2\Phi}| F_3|^2_{\alpha \beta}+ \dots  \;\; = \;\; 0 \;\; ,
\ea\newline
{\bf Background solution.}
We next show that the following ansatz  for the type IIB background metric and dilaton 
\ba\label{Ansatz-BG1}
\Phi &= \phi^{\tiny (0)} +  \alpha \,   \phi^{\tiny (1)}  \;\;\;, \;\;\;\; \phi^{\tiny (0)}  = \text{const.}  \;\; ,\\[0.2 cm]
ds^2_S &= e^{ 2  \alpha \,  A^{\tiny (1)}  }  \eta_{\mu \nu} dx ^\mu dx^\nu + e^{-  2 \alpha  A^{\tiny (1)}  }   \Big( g^{\tiny (0)}_{ij}   + \alpha \,   \, g^{\tiny (1)}_{ij} \Big) dy^i d y^j \;\; ,\label{Ansatz-BG2}
\ea
is a solution to \eqref{EOMdilaton}-\eqref{EinsteinTangent}, where  $g^{\tiny (0)}_{ij}$ denotes the unmodified Calabi-Yau metric which constitutes a solution to the  classical E.O.M.'s. Moreover  $g^{\tiny (1)}_{ij}$ can be expressed in terms of tangent and normal indices $i \to (\alpha,a)$ according to \eqref{tangent_normal_frame} and \eqref{tangent_normal_frame_Inv}. Firstly, one infers from \eqref{Ansatz-BG1} and \eqref{Ansatz-BG2} that  \eqref{EOMdilaton} may be re-expressed as\footnote{Let us emphasize that in this work for simplicity i.e.\,the one-modulus $h^{1,1}=1$ Calabi-Yau  \eqref{Ansatz-BG1}  and \eqref{Ansatz-BG2} are given for the case where the world-volume of 8 D7's and  O7 coincides.  However, note that the \eqref{Ansatz-BG1} and \eqref{Ansatz-BG2} can be generalized to account for non-coinciding D7's and O7's by simple summing over the respective contributions. }
\ba\label{EOMDil2}
  &\nabla^{\tiny(0)} {}^{\alpha} \nabla^{\tiny(0)}{}^{\beta} \Big(g^{\tiny (1)}_{\alpha \beta }  +2 \,\delta_{D7}\, e^{\phi^{\tiny (0)} }\cdot R_T{}_{\alpha \beta}|_{oY_3}  \Big) +    \nabla^{\tiny(0)}{}^{a}\nabla^{\tiny(0)}{}^{b} \Big( g^{\tiny (1)}_{a b } -2\,\delta_{D7}\, e^{\phi^{\tiny (0)}}\cdot \bar R_{ab} |_{oY_3} \Big) \\ \nonumber
  &+   \nabla^{\tiny(0)} _i  \nabla^{\tiny(0)}{} ^i \Big( 4 \, \phi^{\tiny (1)}  -  g^{\tiny (1)}{}_i{}^i + 10 \, A^{\tiny (1)} \big)  -\tfrac{1}{2}| H_3|^2\;\;+\dots\;\; = \;\;0 \;\;.
\ea
We proceed analogously for the Einstein equation \eqref{EinsteinExternal}-\eqref{EinsteinTangent}. For details see appendix \ref{fluxdetails}.
A few comments in order.  From comparison of eq.'s\, \eqref{EOMdilaton} - \eqref{EinsteinExternal}  one  infers that there exists no solution for vanishing background fluxes $H_3$ and $F_3$. To solve for $H_3$ and $F_3$ explicitly is beyond the scope of this work as it additionally requires to check consistency with the  E.O.M's of $H_3, F_3$ and $C_4$.  For our purpose it is sufficient  to limit ourselves to making an Ansatz for the squared contributions
\beq\label{fluxbackground}
| F_3|^2_{\alpha \beta} , \; | H_3|^2_{\alpha \beta} \;\;  \sim \;\;  \mathcal{O}(\alpha)  \;\; \;\; \text{and} \;\;\;\; | F_3|^2_{a b} ,\; | H_3|^2_{ab} \;\;  \sim  \;\;\mathcal{O}(\alpha) \;\;,
\eeq
rather than $H_3$ and $F_3$ itself.
We use that
\beq
 |H_3|^2 =  g^{\tiny(0)}{}^{\alpha \beta} |H_3|^2_{\alpha \beta}+  \delta^{ab} |H_3|^2_{a b}\;\;,\;\text{and} \;\;\;  |F_3|^2 =  g^{\tiny(0)}{}^{\alpha \beta} |F_3|^2_{\alpha \beta}+  \delta^{ab} |F_3|^2_{a b}\;\; \;.
\eeq
Details of the flux-background  ansatz can be found in the appendix \ref{fluxdetails}.
Let us take a step back to discuss the factorization of eq.s\,\eqref{EOMdilaton}-\eqref{EinsteinTangent} in a total derivative contribution and a curvature square density.  One finds that the E.O.M's are of the schematic form\footnote{Note that by commuting two covariant derivatives the two sector in  eq.\eqref{schematicEOM} can communicate in principle.  However, this does not affect our analysis at hand. }
\beq\label{schematicEOM}
\underbrace{\nabla \nabla{}  \Big( \sum_{n }  R_n}_{=\;0} \Big)\;\; + \;\;   \underbrace{ \sum_{m}  R^2_m }_{= \;0}\;\; =\;\; 0  \;\;.
\eeq
where $R_n$ is placeholder for objects in the list  $\left\{ R_T |_{oY_3} , \bar R |_{oY_3}, g^{\tiny (1)},\phi^{\tiny (1)},  A^{\tiny (1)}\right\}$ and $R^2_m$ is our notation for curvature objects in the list $\left\{R_T^2 |_{oY_3},  R^2 |_{oY_3}, \hat R \Omega^2 |_{oY_3}, \Omega^4 |_{oY_3} \right\}$.
Moreover, the formal sum in eq.\,\eqref{schematicEOM} allows for various different index contractions as well as different pre-factors.
 For simplicity, in this work we only provide a solution to the total-derivative contribution which, however suffices to fix \eqref{Ansatz-BG1} and \eqref{Ansatz-BG2}. Note that due to this we can restrict ourselves to making an Ansatz for \eqref{fluxbackground} which only contains total derivative pieces.
\newline
Let us now turn to the discussion of the solution of  \eqref{EOMdilaton}-\eqref{EinsteinTangent}  by  the Ansatz \eqref{Ansatz-BG1} and \eqref{Ansatz-BG2} which is fixed to take the form
\ba\label{metricCorral2}
g^{\tiny (1)}_{\alpha \beta }   \;\;  & =  -2\,\delta_{D7}\,e^{\phi^{\tiny (0)} } \,R_T{}_{\alpha \gamma \beta}{}^\gamma |_{oY_3} + \hat\gamma_3\,\,\delta_{D7}\,g^{\tiny (0)}_{\alpha \beta  }  \cdot e^{\phi^{\tiny (0)} }R_T |_{oY_3} \;\;,  \;\; \\[0.2cm] \nonumber
g^{\tiny (1)}_{a b }   \;\;  & =  \;\; \; 4\,\delta_{D7}\,e^{\phi^{\tiny (0)} }\,\bar R_{a \gamma b}{}^\gamma |_{oY_3} \;\; + \; \hat\gamma_3\,\,\delta_{D7}\,g^{\tiny (0)}_{a b } \cdot e^{\phi^{\tiny (0)} }\,R_T |_{oY_3}  \;\;,
\ea
and moreover the background dilaton and warp factor to be 
\beq\label{eqwarpfactor}
 \phi^{\tiny (1)}    \;\;   =   \big(- \tfrac{6}{5} + \gamma_3 \big)\,\delta_{D7}\,e^{\phi^{\tiny (0)} }\,R_T |_{oY_3}  \;\; , \;\;\;  A^{\tiny (1)}     \;\;   =  \bar\gamma_3 \,e^{\phi^{\tiny (0)} }\,\delta_{D7}\,R_T |_{oY_3} \;,
\eeq
where 
\beq\label{gamma3def}
\gamma_3 = 6\,\hat\gamma_3 +  2 \,\bar\gamma_3\;\;.
\eeq
Thus note that the Einstein equations and dilaton equation of motion alone do not completely fix the  Ansatz \eqref{Ansatz-BG1} and \eqref{Ansatz-BG2}. More precisely, there remains an ambiguity in between the warp factor and the correction to the Calabi-Yau metric parametrized by the $\gamma_3{}'s$. We expect that a complete treatment of the other E.O.M.'s involving the fluxes will fix the remaining freedom.  
\newline
\newline{\bf Ramond Ramond   $C_4$.}
It is interesting to discuss the relationship of the warp-factor \eqref{eqwarpfactor}  and the dynamic equations for the NS-NS and R-R fields given by  \eqref{WarpandF5} and \eqref{BianchiF5}. Counting derivatives one infers that
\beq\label{scalingFluxes}
d F_5 = d \ast_{10} d e^{4 A^{\tiny (1)}} dx^1 \dots dx^4 \sim \mathcal{O} \big(\alpha  \big) \;\;,\;\;  H_3 \sim \mathcal{O} \big(\alpha^{\tfrac{1}{2}} \big) \;\;,\;\;  F_3 \sim \mathcal{O} \big(\alpha^{\tfrac{1}{2}} \big)  \;\;.
\eeq
and thus in particular 
\beq\label{ScalingF5}
F_5 \sim \mathcal{O} \big(\alpha \big) \;\; ,\text{thus} \;\;   |F_5|^2 \sim \mathcal{O} \big(\alpha^{2} \big) \sim  \mathcal{O} {\bf\big(\alpha'{} ^4 \big)} \;\; .
\eeq
This is self-consistent with the E.O.M's  \eqref{EOMdilaton}-\eqref{EinsteinTangent} in which we have  omitted $ |F_5|^2$ due to the fact of being of higher order in $\alpha$.
Moreover, it implies that the Wess-Zumino contribution \eqref{R2-WZ_action} can be safely neglected in the next section \ref{sec:One-mod-red}
as it is  of higher order as well.
Note that \eqref{scalingFluxes} is consistent with the well known flux quantization condition \cite{Giddings:2001yu} given by
\beq
\frac{1}{2 \pi \alpha '} \int H_3  \;\; \in  \;\; 2 \pi \mathbb{Z} \;\; , \;\; \frac{1}{2 \pi \alpha '} \int F_3  \;\; \in  \;\;  2 \pi \mathbb{Z} \;\; .
\eeq
\newline
Let us close this section by providing an outlook on the next section. 
Note that a solution to the equations of motion is a necessary but not a sufficient condition for the background to preserve the required amount of supersymmetry. As we are not aware of a discussion of the  $\alpha'$-corrected supersymmetry conditions we  limited ourselves to the discussion of the E.O.M.'s.
To  gain confidence in the background solution we will provide a check employing four dimensional $\cN=1$ supersymmetry.\footnote{This procedure of fixing higher-derivative terms or for our matter at hand a parameter in the higher-derivative background solution was employed in e.g.\,our previous work \cite{Grimm:2017okk}.  One may  first compactify to lower dimensions and verify  consistency with $4d,\, \cN=1$ or $\cN=2$  supersymmetry. The latter  led us to find  novel higher-derivative terms in type IIA \cite{Grimm:2017okk} which were recently confirmed by scattering amplitudes methods \cite{Liu:2019ses}.}
We use the simple fact that the $\alpha'^2$-correction to  the kinetic terms must take the form \eqref{ActionFromKpot}. In particular, it implies the vanishing of the  $\alpha'^2$-correction to the dilaton kinetic terms as well as the mix terms with the Einstein frame volume $\hat \cV$. Employing this technique in section \ref{sec:One-mod-red} will lead us to fix $\hat\gamma_3$ and $\bar\gamma_3$.

\section{Dimensional reduction  one-modulus Calabi-Yau orientifold}
\label{sec:One-mod-red}

In this section we discuss the  dimensional reduction of \eqref{DBI}, \eqref{ODBI} and \eqref{classicaltypeIIB} on the background solution \eqref{Ansatz-BG1} and \eqref{Ansatz-BG2}.  Let us emphasize that all  equations are treated to linear order in $\alpha$, thus terms of $\mathcal{O}(\alpha^2)$ are neglected systematically. Let us briefly recall some well know features of Calabi Yau orientifold compactifications  of type IIB to four dimensions \cite{Grimm:2004uq, Grimm_2005}. The isometric involution $\sigma$  generated by an $O7$-plane acts on the fields as
\beq
\sigma^\ast \Phi = \Phi \;\;, \;\;\; \sigma^\ast g = g \;\; , \;\; \text{and} \;\; \sigma^\ast  C_4=  C_4 \;\;.
\eeq
The cohomology group  $H^{p,q}$ splits in the even and odd eigen-space  of $\sigma^\ast$ as $H^{p,q} = H^{p,q}_+ \oplus H^{p,q}_-$.
The four-dimensional fields relevant for our discussion arise as\footnote{Where $ \{ \tilde \omega_i\}$ is the basis of $H^{2,2}(oY_3)$. We abuse the notation $i$ for the real coordinates on the internal space, however the meaning should be clear form the context.}
\beq\label{fieldsvrho}
J = v^i \omega_i  \;\;, \;\; C_4 = \rho^i \tilde \omega_i , \;\; i =1\,\dots\, h^{1,1}_ +(oY_3) \;\;.
\eeq
Note that  the range of the index in \eqref{fieldsvrho}  is restricted from the upper bound $h^{1,1}$  in the Calabi-Yau setting to $h^{1,1}_+$ when  orientifold planes are added. In the one-modulus case eq.\,\eqref{fieldsvrho} becomes
\beq
J = \cV^{\tfrac{1}{3}} \,\, \omega  \;\;, \;\;  C_4 = \rho  \,\, \tilde \omega \;\;,
\eeq
where we have used that $ h^{1,1}= h^{1,1}_{+} = 1$.\footnote{In the one-modulus case the single  complex hyper-surface is in the same class as the fix-point locus of the orientifold involution $\sigma(D_m) = D_m$ and thus in the even cohomology. }
As we consider geometric backgrounds with a single K\"ahler  modulus i.e.\,the overall volume,  the scaling of the Calabi-Yau metric is given by
\beq\label{scalingMetric}
 g^{\tiny (0)}_{ij}  \sim \cV ^{\tfrac{1}{3}}  \;\;,
\eeq
 where we abuse our notation as the volume carries dimensions in the background ansatz \eqref{scalingMetric}. It will be cast dimensionless when dressing it with the appropriate $\alpha'$-powers eq.\,\eqref{dimlessVolume} after the dimensional reduction.
 One is next interested in inferring the scaling behavior of the corrections to the background \eqref{metricCorral2} and \eqref{eqwarpfactor}  under eq.\,\eqref{scalingMetric}. Using that the internal space Riemann tensor with  downstairs indices scales as $ \cV^{1/3}$ one concludes that
the higher-derivative corrections to the background   \eqref{metricCorral2} and \eqref{eqwarpfactor} scale  as\footnote{Where we  use that $\partial_\alpha Y ^M \sim \cV^{0}$ and $\xi^M_a \sim \cV^{-1/6}$.}
\beq\label{scalingV}
\phi^{\tiny(1)} \propto \cV^{-\tfrac{1}{3}} \;\;\ ,\;\;\; A^{\tiny(1)} \propto \cV^{-\tfrac{1}{3}} \;\; ,\;\;\; g^{\tiny (1)}_{ij}  \propto \cV^{0} \;\;.
\eeq
Separating the volume modulus dependence  in the background ansatz as  derived in eq.\,\eqref{scalingV}  one finds that 
\ba\label{AnsatzN-BG1}
\Phi &= \phi+  \alpha \, \cV ^{-1/3}   e^{\phi} \phi^{\tiny (1)}  \;\;\;,\\[0.3 cm]
ds^2_S &= e^{ 2  \alpha \, \cV ^{-1/3}    e^{\phi}  \, A^{\tiny (1)}  }  g_{\mu \nu} dx ^\mu dx^\nu + e^{ -2  \alpha \, \cV ^{-1/3}   e^{\phi} \, A^{\tiny (1)}  } \cV ^{1/3}   \Big( g^{\tiny (0)}_{ij}   + \alpha \, \cV ^{-1/3}  e^{\phi} \, g^{\tiny (1)}_{ij} \Big) dy^i d y^j  \;\; , \nonumber
\ea
where  $\phi = \phi(x)$, $\cV = \cV(x)$ and $ g_{\mu \nu} =g_{\mu \nu} (x)$ the  dynamic  external metric.  There is no contribution from the classical DBI and WZ actions to the kinetic terms of the discussed fields \cite{Jockers_2005}. Thus we can omit them from the study at hand. 
Dimensionally reducing the DBI and ODBI action of the coincident  eight  D7's and the single O7 \eqref{DBI} and \eqref{ODBI} one finds  by using eq.\,\eqref{deformationRiemann} that
\ba\label{reduction-DBI_ODBI}
8S_{DBI}^{R^2}  + S_{ODBI}^{R^2}  \longrightarrow \frac{\alpha}{(2 \pi)^5 \alpha'}\int 
& \left( \frac{4 }{9 \, \hat\cV^{8/3}}    \int_{D_m} R_T |_{oY_3}\ast 1 \right)  d \hat\cV \wedge \ast d \hat\cV \;\; \\[0.2 cm] \nonumber
 + & \left(   \frac{ 3}{ \hat\cV^{2/3}}   \int_{D_m} R_T |_{oY_3} \ast 1 \right)    d \phi \wedge \ast d \phi  \; \\[0,2cm]\nonumber
 +&\left(  \frac{8}{3 \, \hat\cV^{5/3}}    \int_{D_m} R_T |_{oY_3} \ast 1\,  \right)   d \phi \wedge \ast d \hat \cV \;\;  + \mathcal{O}(\alpha^2)\;\;,
\ea
where we  have absorbed $(2 \pi \alpha')^3$ to render the volume dimensionless $\cV \to  \cV / (2 \pi \alpha')^3$ which leads  to
\beq
\alpha  \to   \frac{1}{16}  \;\;. 
\vspace{0.2cm}
\eeq
Note that the reduction computation to arrive at \eqref{reduction-DBI_ODBI} exclusively depends on the zeroth order Calabi-Yau background as the $\mathcal{O}(\alpha)$-corrections to the background \eqref{AnsatzN-BG1} lead to $\alpha^2$-contributions and are thus to be neglected. Moreover, to arrive at \eqref{reduction-DBI_ODBI} we made use of   \eqref{DBI-eqs} in combination with \eqref{CYO3-1}-\eqref{CYO3-3} which allowed us to connect total space Riemann curvature components to tangent and normal space curvature expressions.

Let us next present the reduction result in the Einstein frame after a Weyl rescaling\footnote{See eq.\,\eqref{sec_Weylrescaling}.} $g'_{\mu \nu }\;\; = \;\; \Lambda g_{\mu \nu}  $ of the four-dimensional metric by  
\ba\label{Weyl-rescailng-alphaprime}
 \Lambda^{-1}  = e^{-2\phi}\cV + \Big(\tfrac{1}{2} g^{\tiny (1)}_i{}^i - 4 A^{\tiny (1)} - 2 \phi^{\tiny (1)} \Big) \;\;,
\ea
where\footnote{We use that $g^{\tiny (0)}{}^{\alpha \beta}   R_T{}_{\alpha \gamma \beta}{}^\gamma |_{oY_3} =  R_T |_{oY_3} $ and moreover  that  $\delta^{ab}  \bar R{}_{a \gamma b}{}^\gamma |_{oY_3} = - R_T |_{oY_3} $ which follows from eq.\,\eqref{RicciFlatConnection}. }
\ba
g^{\tiny (1)}_i{}^i  \;\; \sim  \;\;A^{\tiny (1)} \;\; \sim  \;\; \phi^{\tiny (1)}  \;\; \sim  \;\; e^{-\phi}\cV^{1/3}  \int_{D_m} R_T |_{oY_3} \ast 1 \;\;.
\ea
Note that \eqref{Weyl-rescailng-alphaprime} implies that the reduction result before the Weyl-rescaling contains the gravitational term
\beq\label{EHcorr4d}
\frac{1}{(2 \pi)^4 \alpha'}  \cdot  \Lambda^{-1}  \int  
R \ast_4 1  \;\; .
\eeq
By using eq.\,\eqref{deformationRiemann}  and eq.'s\,\eqref{tangent_normal_frame}-\eqref{RicciFlatConnection2} one infers that the reduction of the classical action \eqref{classicaltypeIIB}    results in
\ba\label{reduction-classical0}\nonumber
  S^0_{IIB - S}  \longrightarrow \frac{1}{(2 \pi)^4 \alpha'} \int 
R \ast_4 1 \;\; &+ \;   \left(-\frac{2}{3  \hat\cV^{2}} - \frac{\tfrac{388}{5}\, \alpha }{9 \, \hat\cV^{8/3}}  \int_{D_m} \tfrac{1}{2 \pi} R_T |_{oY_3}\ast 1 \right)  d \hat\cV \wedge \ast d \hat\cV \;\; \\[0.2 cm] \nonumber
 + & \; \left(-\frac{1}{2} -  \frac{ 3 \alpha}{ \hat\cV^{2/3}} \int_{D_m} \tfrac{1}{2 \pi}R_T |_{oY_3}\ast 1 \right)    d \phi \wedge \ast d \phi  \; \\[0,2cm]
 +& \;\left( - \frac{8 \, \alpha}{3 \, \hat\cV^{5/3}}  \int_{D_m} \tfrac{1}{2 \pi}R_T |_{oY_3} \ast 1\,  \right)   d \phi \wedge \ast d \hat \cV \;\;  + \mathcal{O}(\alpha^2).
\ea
To make \eqref{reduction-DBI_ODBI} and \eqref{reduction-classical0}  compatible with four-dimensional $\cN = 1$ supersymmetry following \eqref{ActionFromKpot} we fixed  the two free parameters in the background solution   \eqref{metricCorral2} -\eqref{gamma3def} to be 
\beq\label{fix_factorgamma3}
\hat\gamma_3 = -\tfrac{71}{18} \;\; ,\;\;  \bar\gamma_3 =\tfrac{113}{15}\;\; \to \;\; \gamma_3 = -\tfrac{43}{5} \;\; . 
\eeq
Note that this relies on the assumption that the imaginary part of the axio-dilaton eq.\,\eqref{axio-dilaton} does not receive corrections, see section \ref{theObjective} for details.
Combining \eqref{reduction-DBI_ODBI} and \eqref{reduction-classical0} one thus infers that
\ba\label{totalreddvdv}
  &S^0_{IIB - S} + 8S_{DBI}^{R^2}  + S_{ODBI}^{R^2}  \longrightarrow  \\[0.2 cm] \nonumber
 \frac{1}{(2 \pi)^4 \alpha'} \int R \ast 1 & -\frac{1}{2}  d \phi \wedge \ast d \phi -\left(\frac{2}{3  \hat\cV^{2}} + \frac{\tfrac{368}{5} \cdot \alpha }{9 \, \hat\cV^{8/3}} \int_{D_m} \tfrac{1}{2 \pi} R_T |_{oY_3} \ast 1\,\right)  d \hat\cV \wedge \ast d \hat\cV \, ,
\vspace{0.2cm} \ea
 where in \eqref{reduction-classical0} and \eqref{totalreddvdv} we have neglected the kinetic terms of the fields arising from $H_3$ and $ F_3$  since those are irrelevant for our analysis. The contribution from  $F_5$ will be discussed below in the text. Moreover, we have assumed that the $R^2$-action of eight coinciding $D7$-branes is simply eight times $S_{DBI}^{R^2}$. 
A few comments in order. Firstly, note that the order $\alpha$-contributions to \eqref{reduction-classical0} arise exclusively from the corrected background \eqref{Ansatz-BG1} and \eqref{Ansatz-BG2}. 
Secondly, note that  $\cN=1$ supersymmetry \eqref{ActionFromKpot} requires the $\alpha'^2 g_s$-correction  of the  mix kinetic terms of the dilaton and the Einstein frame volume as well as  the dilaton kinetic terms to vanish. 
 The  caveat to this approach is that turning on gauge flux may alter the E.O.M's and thus the background solution. Note that we did not solve all the E.O.M's of the system and thus a solution with vanishing D-brane gauge field flux may be inconsistent.  However, note that  the integrated condition \eqref{d3tadpoleMod} is consistent with vanishing gauge flux.
   \newline
{\bf{$|F_5|^2 $-terms.}} 
Let us next turn to the scalar field $\rho$ which arises in the reduction of the Ramond-Ramond five-form field strength.
The $C_4$ four-form field gives rise to a scalar as
\beq
C_4 \;\; =  \;\; \rho  \cdot \tilde \omega
\eeq
where $\rho = \rho(x)$ and $ \tilde \omega$ is the unique harmonic four-form. The  $\alpha$-contribution  to the  $\rho$-kinetic terms arising from the classical action \eqref{classicaltypeIIB}  origin solely from the $\alpha$-correction to the background metric \eqref{AnsatzN-BG1}. After a Weyl rescaling by \eqref{Weyl-rescailng-alphaprime} to the four-dimensional Einstein frame  and by using \eqref{fix_factorgamma3} one finds
\ba\label{F52-reductionResult}
\tfrac{1}{2 \kappa_{10}} \int  -\tfrac{1}{4} |F_5|^2 \ast_{10 } 1  \rightarrow \tfrac{1}{(2 \pi)^4 \alpha'} \int 
  &  -\left(   \frac{1}{6 \,  \cV^{ 4/3}} +  \frac{\tfrac{374}{5}\cdot \alpha  }{ 36\, \cV^{2}}  \int_{D_m} \tfrac{1}{2\pi}R_T |_{oY_3} \ast 1  \right) e^{2 \phi}d \rho \wedge \ast d \rho \, ,
\ea
where we have used that the divisor is minimal i.e.\,K\"ahler for the metric \eqref{metricCorral2}. Note that the left hand-side in eq.\,\eqref{F52-reductionResult} is part of the  ten-dimensional type IIB supergravity action in the string frame.
Let us emphasize  that the higher-derivative $F_5^2 R$-terms in the DBI and ODBI action remain elusive.
One may next use
\beq\label{RTasZ}
\int_{D_m} R_T |_{oY_3} \ast 1  =  4 \pi \cZ \;\; ,
 \eeq
to express the total reduction result in  terms of manifestly topological quantities.  The complete four-dimensional action is the sum of  \eqref{reduction-DBI_ODBI},\eqref{reduction-classical0} and \eqref{F52-reductionResult} and results in  
\ba\label{Result-Action}
S = \frac{1}{2 \kappa_4^2} \int R \ast 1 \;\;
 - \frac{1}{2}  d \phi \wedge \ast d \phi 
\; - & \left( \frac{2}{3 \,\hat\cV^{2}  } + 2 \,\alpha \cdot  \frac{ \tiny 368 /5   }{9 \hat\cV^{8/3}}  \; \cZ   \right) d \hat\cV \wedge \ast d \hat\cV \;\; \\[0.2 cm] \nonumber
-   &  \left( \frac{1}{6 \, \hat \cV^{ 4/3}} + 2 \, \alpha \cdot  \frac{  \tiny 374/5+\gamma_4 }{36\, \hat\cV^{2}}  \; \cZ  \right)d \rho \wedge \ast d \rho  \;\; .
\ea
With $2 \kappa_4^2 = (2 \pi)^4 \alpha'$ and where we have introduced a new parameter which is expected to obey
\beq
\gamma_4 \;\; \neq  \;\;  0\;\;,
\eeq
to highlight the fact that the DBI and ODBI $F_5^2 R$-terms  are currently unknown. However, the latter are expected to contribute to the $\rho$-kinetic terms in \eqref{Result-Action}.
\newline
Finally, we are in a position to draw conclusions on the K\"ahler potential and coordinates.
By comparing  eq.\,\eqref{ActionFromKpot}  with eq.\,\eqref{Result-Action} one fixes the parameters  in the Ansatz \eqref{Kpot1M} and \eqref{Kccord1M}   for the K\"ahler potential and coordinates as
\ba\label{gam1gam2Result}
\gamma_1 =  \frac{1}{128} \cdot \big( \gamma_4 - 146 \big) \;\;\; \text{and} \;\;\; \gamma_2 = -\frac{1}{48}   \cdot \big( \gamma_4 + \tfrac{6}{5} \big)   \;\;.
\ea
Let us close this section with a few brief comments. Firstly, note that eq.\,\eqref{gam1gam2Result} implies that for any value of $\gamma_4$ either the correction to the K\"ahler potential or coordinates is present. For $\gamma_4 \neq - 6/5$  one in particular finds that the  no-scale structure is broken by the $\alpha'^2 g_s$-correction. We will discuss an indirect way to fix  $\gamma_1$  and  $\gamma_2$ in section \ref{FtheoryMatch}.
\newline
{\bf Comments.}  Let us close this section with a short discussion on the presence of the Einstein-Hilbert term in the reduction result eq.\,\eqref{EHcorr4d}.\footnote{The one-loop case has been discussed in \cite{Haack:2015pbv}. I would like to thank Michael Haack for his helpful comments in particular  on this topic.} Note that by using eq.\,\eqref{EHcorr4d}  as well as eq.\,\eqref{RTasZ} we find a correction to the Einstein-Hilbert term as
\beq\label{EHcorrZ}
\sim \;\; \frac{1}{(2 \pi)^4 \alpha'}  \int R \ast 1  \, \Big( \cV^{\tfrac{1}{3}} \, e^{-\phi} \, \mathcal{Z} \Big) \;\; ,
\eeq
where the volume is dimensionless. Firstly, note that one may easily infer  that this is of order $\alpha'^2 g_s$ relative  to the leading order term which in the string frame scales as  $\sim  \cV  \, e^{-2\phi} $. Secondly, it is absent for four-cycles with vanishing first Chern form  thus in particular for flat backgrounds which follows from eq.\,\eqref{EHcorrZ}. The $R^2$-sector  fixed by the open string disk and projective plane amplitudes  \cite{Bachas:1999um}  upon reduction does not give rise to  an Einstein-Hilbert term correction to the four-dimensional theory. One thus expects latter to arise by matching effective field theory with the amplitudes of closed string gravitons scattered off D-branes at disk level. However, such terms have not been identified at the two-point level \cite{Garousi_1996,Hashimoto_1997}.  Note that usually the gravitons are scattered off flat D-branes in which case  the correction eq.\,\eqref{EHcorrZ} is trivially absent. It is of interest to conduct this study for higher-point functions, in particular for D-brane world-volumes with non-vanishing intrinsic Ricci curvature.

\section{Discussion of results and conclusions}

\subsection{Connection to F-theory and the generic moduli case} \label{FtheoryMatch}
In this section we focus on two main points. Firstly, we show that the form of the topological $\cZ$-correction \eqref{Z-def} can be matched with our previous F-theory approach \cite{Weissenbacher:2019mef}, in particular with the F-theory setting admitting non-Abelian gauge groups \cite{Grimm:2013bha}. Secondly,  note that the K\"ahler coordinates during the F-theory uplift are expected to receive one-loop effects \cite{Tong:2014era,Weissenbacher:2019mef,Grimm:2017pid}, which constitutes the main obstacle to performing a conclusive F-theory analysis.  Although the K\"ahler-potential  in the F-theory lift  may as well potentially be corrected at loop-level the "$\alpha'^2$-corrections" to it can formally be up-lifted by using the classical formalism \cite{Grimm_2011,Grimm:2013bha,Weissenbacher:2019mef}. One may thus match it with the type IIB approach taken in this work in particular eq.\,\eqref{Kpot1M}. This will allow us  to suggest  values for  $\gamma_1$ and $\gamma_2$.

F-theory incorporates the physics of D7-branes and O7-planes in the geometry of an elliptically fibered Calabi-Yau fourfold $Y_4$ with K\"ahler base $B_3$  \cite{Vafa_1996,weig2018tasi,donagi2008model,Grimm_2011}. 
In particular, the weak-coupling limit of F-theory is equivalent to type IIB compactified on a Calabi-Yau orientifold $oY_3$, i.e.\,a Calabi-Yau background $Y_3$  with orientifold planes added. One obtains the manifold $Y_3$ by taking the double cover of the base $B_3$  branched along the
orientifold locus, i.e.\,the four-cycle wrapped by the O7-plane. One thus  identifies
\beq
B_3 \equiv oY_3 \;\; .
\eeq
Let us start by giving some generic results relevant for the discussion which follows in this section. For a single O7-plane with locus $D_m \subset B_3 $  one finds that\footnote{ We use the abbreviation $[\omega] = [PD(\omega)]$, where $\omega$ is a $(1,1)$-form.}
\beq\label{relc11}
 [c_1(B_3)] = \tfrac{1}{2}[D_m]  \;\; ,
\eeq
and by using adjunction that
\beq\label{relc12}
[c_1(D_m)] = - [c_1(B_3)]  \;\; ,
\eeq
where $c_1(B_3)$ and $c_1(D_m)$ are the first Chern forms of the base and the orientifold locus, respectively.
The  four-dimensional K\"ahler potential in the weak coupling limit of F-theory  \cite{Weissenbacher:2019mef,Grimm:2013bha} takes the form\footnote{Where we omit the potential novel corrections  \cite{Weissenbacher:2019mef} to the K\"ahler potential which are not of the form \eqref{FtheoryCorrection},\eqref{FtheoryCorrection2} and \eqref{FtheoryCorrection3} .}
\beq\label{FtheoryCorrection}
K_{ F-theory} =  \phi  - 2\, \text{log}\Big( \hat\cV +  \tfrac{1}{96} \hat\cV^{\tfrac{1}{3}} \cZ_{ F}  \,\Big) \;\; ,
\eeq
where here we have used that  the base manifold  admits a single K\"ahler modulus and that
 \beq\label{FtheoryCorrection2}
 \mathcal{Z}_{ F} =   \int_{B_3} \, PD(\mathcal{C})\, \wedge \omega \;\;,    \;\;\;\;  \mathcal{C} \subset B_3 \;\; .
\eeq
where $\omega $ is the $(1,1)$-form  Poincare dual to the single complex co-dimension one hyper-surface in $B_3$. The complex curve $\mathcal{C} $ and its Poincare dual four-form $PD(\mathcal{C})$ are fixed by matching it to the corresponding correction in the three-dimensional $\cN=2$ theory. In particular, the curve $\mathcal{C}$  is determined by  the F-theory lift of the three-dimensional K\"ahler potential. One finds \cite{Weissenbacher:2019mef,Grimm:2013bha}  that by shrinking the torus inside the elliptically fibration $T^2 \to 0$   and by moreover taking the weak coupling limit \cite{Sen_1996}  that the curve  $\mathcal{C}$ is fixed by the Calabi-Yau fourfold information  as
\beq\label{FtheoryCorrection3}
 \mathcal{Z}_{ F} \stackrel{ !}{=} \lim_{\tiny w.\,c.} \cZ_{3d} \; , \;\;  \text{with} \;\;\;   \cZ_{3d}= \int_{Y_4}  c_3(Y_4) \wedge \omega \;\;.
\eeq
 Here $ c_3(Y_4) $ is  the third Chern-form of the Calabi-Yau fourfold. 
In  \cite{Grimm:2013bha} an extended study of F-theory backgrounds admitting $n$-stacks   with $SU(N_i), \; i=1,\dots,n$ gauge groups was conducted. Leading to  a total gauge group of
\beq
G = \prod_i^{n} SU(N_i) \;\; .
\eeq 
This led us to suggest\footnote{We use the notation $[D_1] \cdot [D_2]$ to denote the intersection product between two sub-varieties $[D_1]$ and $[D_2]$.}
\beq\label{conjectureC}
 \mathcal{C} =  - [W] \cdot \Big( [W] -\tfrac{1}{2}[c_1]  \Big) - \sum_{\sigma = \pm} \sum_{i=1}^{n} N_i \, [S^{\sigma}_{i}] \cdot \Big( [S^{\sigma}_{i}] + \tfrac{1}{2}[c_1]  \Big) \;\;.
\eeq
Where  $[W]$ is the class of the Whitney umbrella \cite{Braun_2008,Collinucci_2009},  $[S^{\pm}_{i}]$ are the hyper-surfaces wrapped by the $i^{th}$-brane stack and its orientifold image and
\beq
[c_1]  = [\pi^\ast c_1(B_3)] \;, \;\; \text{with} \;\;\;\; \pi: Y_3 \to B_3 \;\; ,
\eeq
is the first Chern form of the base  pulled back by the projection from the double cover $Y_3$ to the base $B_3$.
On the type IIB orientifold \cite{Brunner_2004,Blumenhagen_2007,denef2008les} with eight coincident D7-branes and one $O7^-$-plane the gauge group  is given by
\beq\label{gaugetypeIIB}
G_{8D7+ O7} = SO(8)  \;\;.
\eeq
Let us next turn to eq.\,\eqref{conjectureC} in the one-modulus case with a single divisor class $[D_m]$ of the base  and a  $SU(N)$ gauge group.  One finds  by using  \eqref{relc11} that
\beq\label{curveRed}
 \mathcal{C}  =  - \tfrac{1}{4} [D_m] \cdot  [D_m]  \, \big( 3 + 10\, N\big)\;\;.
\eeq
where we have used that $[S^{\pm}] = [W] = [D_m] $,  and eq.'s \eqref{conjectureC} -\eqref{gaugetypeIIB}. The relations  \eqref{relc11}  and \eqref{relc12} hold both expressed as classes in the base i.e.\,in $oY_3$ as well as for the double cover $Y_3$. Thus for notational simplicity we omit a distinction in this section.
Finally, by using  adjunction one finds eq.\,\eqref{relc12}  and thus from  eq.\,\eqref{relc11}  that
\beq
 [c_1(D_m)] =  -\tfrac{1}{2}[D_m] \;\; , 
 \eeq
which leads us to arrive at\footnote{Note that $[\omega]  = [D_m]$. We chose to express \eqref{ZfZrel} with explicit appearance of $\omega$ as it is closer to the schematic form of the correction in the generic moduli case. } 
\beq\label{ZfZrel}
 \mathcal{Z}_{ F}   \sim     \mathcal{Z} =  -\tfrac{1}{2}\, [D_m] \cdot [D_m] \cdot [\omega] \;\;.
\eeq
As the study performed in \cite{Grimm:2013bha}  and thus relation \eqref{conjectureC} does not incorporate for $SO(8)$ gauge groups eq.\,\eqref{ZfZrel} constitutes a heuristic argument. 
However, with that caveat in mind we conclude that the correction $ \mathcal{Z}_{ F} $ derived in F-theory and the $\mathcal{Z} $-correction of the type IIB approach are of the same topological form.
 It would be interesting to study eq.\eqref{FtheoryCorrection2}  in F-theory setups which admit $SO(8)$ gauge groups. 
 Alternatively, one may turn on gauge flux in the type IIB setting which will lead to different gauge groups such as
\ba\label{gaugeGroups}
G_{6  D7 + {\bf D7}+ {\bf D7'}+ O7} &= SO(6) \times U(1) \;\;, \;\; \nonumber  
G_{4  D7 + 2{\bf D7}+  2{\bf D7'}+ O7} = SO(4) \times U(2) \;\;, &\\ 
G_{2  D7 + 3{\bf D7}+  3{\bf D7'}+ O7} &= SO(2) \times U(3) \;\;, \;\;
G_{ 4{\bf D7}+  4{\bf D7'}+ O7}  =   U(4)\,,  &
\ea
where the bold notation refers to the type IIB  seven branes which are shifted slightly away from the O7-plane due to the gauge flux.\footnote{The branes ${\bf D7}$ and  $ D7$ lie in the same Homology class.} On the F-theory side \cite{Grimm:2013bha}  one finds that for $U(1)$-restricted models   for simple non-Abelian gauge groups such as $SU(N)$ the relation \eqref{conjectureC} results in
\beq\label{conjectureC2}
 \mathcal{C} =  \sum_{\sigma = \pm} \left(- [W^{\sigma}]\cdot \Big([ W^{\sigma}]-\tfrac{1}{2} [c_1] \Big) -  \sum_{i=1}^{n} N_i  \, [S^{\sigma}_{i}] \cdot \Big([S^{\sigma}_{i}] + \tfrac{1}{2} [c_1]  \Big) \right) \;\;,
\eeq
where the Whitney umbrella is shifted away from the orientifold locus such that it additionally contributes its image thus $ W^{\pm}$.
Let us next establish a connection to the DBI type IIB side. Under the assumption that a non-vanishing gauge flux does not alter the  type IIB discussion of the K\"ahler metric  one may now match the K\"ahler potential on the F-theory side with the one  derived from the DBI actions. Note that since 
\beq
SU(4) \simeq SO(6) \;\;,
\eeq
 the type IIB setup eq.\eqref{gaugeGroups} with seven D7's and one O7 can be matched with our formula \eqref{conjectureC2} for $U(1)$ restricted models in F-theory.
In that case one encounters
\beq
 \mathcal{C}  =  - \tfrac{23}{2} \; [D_m] \cdot [D_m]  \;\; .
\eeq
Moreover, note that on the type IIB side we need to change the number of D7-branes and thus our DBI action pre-factor becomes 
\beq
\alpha \to \frac{11}{192}\;\;.
\eeq
  We are now in a position to identify   the   K\"ahler potential  obtained via  F-theory eq.\,\eqref{FtheoryCorrection} to the one obtained from the  DBI actions of  seven D7-branes and a single O7-plane. 
By using eq.\,\eqref{FtheoryCorrection} one can fix the K\"ahler potential eq.\,\eqref{Kpot1M}  on the type IIB side  to
\beq\label{eqfinalgammas}
\gamma_1 =   \frac{23}{96}  \;\;  \Rightarrow \;\; \gamma_2 =  -\frac{69}{20} \;\; ,
\eeq
which allows us to derive the K\"ahler coordinate correction $\gamma_2 $.

Concludingly, the  comparison to F-theory  suggests that the no-scale  structure is broken  due to the $\alpha'^2 g_s$-correction to the K\"ahler-coordinate \eqref{Kccord1M} and \eqref{eqfinalgammas}. Note that the sign of $\gamma_2$ agrees with our F-theory discussion in setups without non-Abelian gauge groups  \cite{Weissenbacher:2019mef}. In the presence of a non-vanishing flux-superpotential in the vacuum one may use this correction to stabilize the K\"ahler moduli, in AdS, Minkowski as well as  potentially de Sitter vacua as discussed in \cite{Weissenbacher:2019bfb}.

Let us close this section with some remarks on the generic K\"ahler moduli case of  the Calabi-Yau orientifold. While the dimensional  reduction of the $\alpha'^2 $-corrected DBI action of the D7-branes and O7-planes is performed in the one-modulus case one may use the F-theory side of the derivation to gain confidence in eq.'s \eqref{Kpot}  and \eqref{Kccord}. In particular, eq.'s \eqref{FtheoryCorrection2}  and \eqref{ZfZrel}  suggest that
\beq
\cZ_i  \sim  \int_{oY_3} \, \mathcal{C}\, \wedge \omega _i \;\;,
\eeq
where $i = 1,\dots, h^{1,1}_+$ of the Calabi-Yau orientifold and $\omega_i$ the harmonic $(1,1)$-forms and the curve $\mathcal{C}$ in \eqref{conjectureC} and \eqref{conjectureC2}, respectively.
\vspace{0,2cm}
\newline
{\bf D3 brane instantons.} Let us close this section by discussing the $\alpha'^2 g_s$-correction to the K\"ahler coordinates  \eqref{Kccord1M}. As the K\"ahler coordinates linearize the Euclidean D3-brane action the $\mathcal{Z}\log(\cV)$-correction admits an interpretation as a  loop effect. Potentially originating from the one-loop determinant.
  Moreover, the study of its relation to the conformal anomaly of a $4d,\, \cN=1$ SCFT is of interest \cite{ArkaniHamed:1997mj,Bobev:2013vta}. This may also suggest a connection to the correction derived in \cite{Conlon_2010}.  In latter the authors suggest a correction of the K\"ahler coordinates in the low-energy limit as
\beq\label{betaT}
T = 3 \hat\cV^{\tfrac{2}{3}} +  \frac{\beta }{12 \pi} \, \text{log} \hat\cV
\eeq
 where $\beta$ is the beta-function  underlying the running of the gauge coupling. This discussion originated from the study of threshold corrections to the gauge couplings of branes at orientifold singularities  in local models \cite{Conlon_2009}. Firstly, let us note that both the corrections in eq.\,\eqref{Kccord1M} as well as in eq.\,\eqref{betaT} are of order $\alpha'^2 g_s$. 
Moreover, $\beta$  as well as  $\cZ$  only depend on characteristics of the gauge group. Based  on this heuristic comparison  one may  suggest that
  \beq
 \beta \;\;  \sim  \;\;  \int_{oY_3} \, \mathcal{C}\, \wedge \omega \;\; ,
 \eeq 
  where $\omega$ is the $(1,1)$-form Poincare dual to the hyper-surface wrapped by the D-branes and $\mathcal{C}$ is the curve given in eq.\,\eqref{conjectureC} and  eq.\,\eqref{conjectureC2}, respectively. It would be of great interest to study this potential alternative origin of our $\alpha'^2 g_s$-correction eq.\,\eqref{Kccord1M} to gain a better physical understanding. 
  Lastly, let us emphasize that  an analogous discussion may be carried out related to  our previous result in $3d,\,\cN=2$ supergravity  \cite{Grimm:2017pid}, where the K\"ahler coordinates resemble the linearized M5-brane action. 

\subsection{On $\alpha'$-corrections from D$p$-branes} \label{D5branes}
The main part of this work studied higher-derivative corrections stemming from D7-branes and O7-planes. One may analogously study the effects of other space-time filling D-branes and O-planes  in type IIB  - and type IIA - to the K\"ahler potential of the $4d,\,\cN=1$ theories.   D$p$-branes and O$p$-planes admit the  leading order correction to the DBI action  eq.\,\eqref{DBI}  and  eq.\,\eqref{ODBI} but with brane tension
\beq\label{muDpbrane}
\mu_p \;\; =  \;\; (2 \pi)^{-p} \, \alpha'^{-\tfrac{p+1}{2}} \;\;,
\eeq
and with  $p$-dimensional brane world-volume.

{\bf $\alpha'^3 g_s$-corrections from D5-branes.}
Let us start by analyzing space-time filling D3-branes and O3-planes which are localized as points in the internal geometry and thus the leading order corrections to the DBI action do not contribute to the kinetic terms upon dimensional reduction. Space-time filling  D9-branes and O9-planes in principle may generate corrections to the kinetic term of the volume modulus proportional to the Ricci-scalar of the tangent directions of the brane in the internal space. However, the  latter is vanishing for Calabi-Yau backgrounds.

Let us next turn to a more detailed discussion of space-time filling D5-branes and O5-planes. Those wrap a holomorphic 2-cycle  $ C_m$ inside the Calabi-Yau orientifold i.e.\,the minimal 2-cycle inside the Homology class. We refrain from  discussing tadpole constraints and deriving the E.O.M.'s for this system here, but instead focus  on the dimensional reduction of  the action \eqref{DBI}  with $p=5$ in eq.\,\eqref{muDpbrane}
on the one-modulus Calabi-Yau background of the form
\ba\label{AnsatzBGD5}
ds^2_S &= g_{\mu \nu}  dx ^\mu dx^\nu +  \cV ^{1/3}   g^{\tiny (0)}_{ij} dy^i d y^j  \;\; , 
\ea
where $g^{\tiny (0)}_{ij} $ is the Calabi-Yau metric,  $g_{\mu \nu} = g_{\mu \nu} (x)$ the external metric and $\cV  = \cV (x)$ the overall volume modulus. Thus with  $\mu_5 = ( (2 \pi)^2 \cdot (2 \pi \alpha')^{3})^{-1} $ and by using eq.\,\eqref{deformationRiemann} one infers that the leading order correction to the D5-brane DBI action results in
\ba\label{reduction-DBI_D5}
S_{DBI}^{R^2}  \longrightarrow \frac{\tilde\alpha}{(2 \pi)^4 \alpha'}\int 
& \left( \frac{4 }{9 \, \hat\cV^{3}} \, e^{-\tfrac{\phi}{2}} \mathcal{X}  \right)  d \hat\cV \wedge \ast d \hat\cV 
 +  \left(   \frac{ 3}{ \hat\cV } \, e^{-\tfrac{\phi}{2}} \mathcal{X} \right)    d \phi \wedge \ast d \phi  \; \\[0,2cm]\nonumber
 +& \left(  \frac{8}{3 \, \hat\cV^{2}} \, e^{-\tfrac{\phi}{2}}  \mathcal{X}  \right)   d \phi \wedge \ast d \hat \cV \;\;  + \mathcal{O}(\alpha^2)\;\;,
\ea
where we have defined the topological quantity
\beq\label{topXD5}
\;\;\mathcal{X}  \;= \;  \int_{C_m} R_T |_{oY_3} \ast_2 1  \; =  \; -4 \pi \, \int_{C_m}  c_1(C_m)    \;\;,
\eeq
 with  $\tilde\alpha =   1 / 192$ for a single D5-brane, and with $\hat \cV$ the Einstein frame volume given in eq.\,\eqref{VolumeEinstein}. 
The reduction result \eqref{reduction-DBI_D5} was subject to a Weyl rescaling of the form $g_{\mu\nu} \to e^{2 \phi} \, \cV^{-1} \, g_{\mu\nu}$. Let us emphasize  that \eqref{reduction-DBI_D5} does not constitute the full reduction result as one needs to discuss corrections to the background fields induced by the higher-derivate DBI terms. With that in mind let us note that the corrections to the kinetic terms in \eqref{reduction-DBI_D5} may originate from a novel $\alpha'^3 g_s$-correction to the K\"ahler potential of the form
\beq\label{KpotNew}
K = \phi - 2\, \text{log}\Big(\hat\cV + \gamma_5  \,e^{-\tfrac{\phi}{2}} \mathcal{X} \Big) \;\; ,
\eeq
where $\gamma_5$ is a real number which we do not attempt to fix in this work. A complete study would require the derivation of the $\rho$-kinetic terms from the $F_5^2 R$-sector to determine if a $\alpha'^3 g_s$-correction to the K\"ahler coordinates is present. 
Note that \eqref{KpotNew} is leading order in $g_s$ compared to the well-known Euler characteristic correction \cite{Antoniadis_1997,Becker_2002} and as well breaks the no-scale condition as
\beq
 3 + \gamma_5 \, \frac{3}{2 \hat \cV} e^{- \tfrac{\phi}{2}} \mathcal{X} \;\; .
\eeq
While usually one considers D5-branes and O5-planes simultaneously one may also satisfy the tadpole constraint of a D7/O7 system with additional D5-branes \cite{Plauschinn_2009}. In latter scenario the D5-branes preserve supersymmetry only in special points in the moduli space. However, as such a setup is conceivable in principle one may benefit from the potential correction \eqref{KpotNew}. For instance when stabilizing moduli purely perturbatively where the mechanism relies on the explicit topological numbers of the background \cite{Weissenbacher:2019bfb}. Thus the presence of the correction \eqref{KpotNew} modifies the overall scalar potential and weakens the dependence on the contribution resulting from the Euler-characteristic correction to the K\"ahler potential.
\newline
{\bf D6-branes in type IIA.} 
The intention of this subsection is to initiate the study of $\alpha'$-corrections to the $4d,\,\cN=1$ stemming from D6-branes and O6-planes in type IIA. A  comprehensive study  - as perfomed for the D7-branes in type IIB in the majority of this work - would involve solving the background equations which is beyond the scope of our discussion here. 
We will simply discuss the  dimensional reduction the analog of \eqref{DBI} and \eqref{ODBI} for D6-branes and O6-planes on the background solution \eqref{AnsatzBGD5} for no background fluxes. As the leading order background solution \eqref{AnsatzBGD5} is same for  type IIA and IIB we refrain from rewriting the equation here. In other words the integral in eq.\,\eqref{DBI} is instead to be taken to be over the world-volume of a D6-brane and moreover with brane tension $\mu_6 =  ( (2 \pi)^3 \cdot (2 \pi \alpha')^{3} \cdot  \alpha'^{1/2})^{-1} $.
The Calabi-Yau orientifold  is defined as $oY_3 = Y_3 /  \sigma$, where  $\sigma: Y_3 \to Y_3$ in type IIA string theory is an isometric and anti-holomorphic involution \cite{acharya2002orientifolds,Brunner_2004,Sen_1996}. I.e. \,$\sigma^2 = id$ while preserving the complex structure and metric on infers the action on the K\"ahler form to be
\beq
\sigma^\ast J = J \;\;,
\eeq
where $\sigma^\ast$ denotes the pullback map. For O6-planes on infers that $\sigma^\ast \Omega = - e^{i \theta} \Omega$, with $ \Omega \in H^{(3,0)}$ being the unique holomorphic $(3,0)$-form and $\theta$ a phase angle.  The fix point locus of the involution $\sigma$ is a special Lagrangian three-cycle wrapped by the O6-plane.
The three cycle $D_m$ inside $oY_3$ being special Lagrangian is equivalent is to it being minimal, i.e.\,the representative inside the Homology class which minimizes the volume \cite{Becker_1995}. The D6-branes need to be calibrated w.r.t.\,the same angle $\theta$ as the O6 plane. 
 The isometric involution $\sigma$  generated by an $O6$-plane acts on the fields as
\beq
\sigma^\ast \Phi = \Phi \;\;, \;\;\; \sigma^\ast g = g \;\; , \;\; \text{and} \;\; \sigma^\ast  B_2=  - B_2 \;\;.
\eeq
The cohomology group  $H^{p,g}$ splits in the even and odd eigen-space  of $\sigma^\ast$ as $H^{p,q} = H^{p,q}_+ \oplus H^{p,q}_-$.
We have  that $ h^{1,1}= h^{1,1}_{+} = 1$.
Moreover, the setup-needs to satisfy the tadpole condition
\beq\label{d7tadpoleMod}
\sum_{i}^{\# D6's} N^i_{D6} \cdot \Big(\, [D^{\tiny D6}_i] + [D^{\tiny D6}_i{}'] \,\Big) + 4 \sum_{i}^{\# O6's}  [D^{\tiny O6}_i]  =  0 ,
\eeq
where $[\cdot]$ denotes the class of the four-cycle wrapped by the D6's and O6's and the prime denotes the orientifold image i.e.\,the action on the background geometry in the presence of the O6 plane, see e.g.\,\cite{Plauschinn_2009}.
By dimensionally reducing the DBI and ODBI action of four coincident D6's and on top of a single O6 \eqref{DBI} and \eqref{ODBI} and by a Weyl rescaling  by $g_{\mu\nu} \to e^{2 \phi}\,  \cV^{-1} \, g_{\mu\nu}  $  to the four-dimensional Einstein frame one infers
\ba\label{reduction-DBI_ODBI-D6}
4\, S_{DBI}^{R^2} +S_{ODBI}^{R^2}  \longrightarrow \supset \frac{  \bar\alpha}{(2 \pi)^4 \sqrt{2 \pi}\alpha'}  \int  & \left( \frac{2 }{3\,  \cV^{17/6}}  \,  e^{\phi} \, \int_{D_m}  R_T |_{oY_3}\ast_3 1 \right)  d \cV \wedge \ast d  \cV \;\;  \\[0,2cm]\nonumber
 +& \left( \frac{4 }{3\,  \cV^{11/6}}  \,  e^{\phi} \, \int_{D_m}  R_T |_{oY_3}\ast_3 1 \right)  d \phi \wedge \ast d  \cV
\ea
where we have  absorbed $ (2 \pi \alpha')^3$ in the volume  $ \cV / (2 \pi \alpha')^3 \rightarrow  \cV$ to render it dimensionless  and with
$  \bar\alpha = \frac{1}{32} $ 
accounting for the respective contributions from the four D6-branes and single O6-plane.
Let us emphasize that in particular  in eq.\,\eqref{reduction-DBI_ODBI-D6} no correction to the Einstein-Hilbert term in the four-dimensional theory is induced. Thus the four-dimensional dilaton is un-modified. Moreover we omit terms which carry external derivatives of the dilaton in \eqref{reduction-DBI_ODBI}.
Note that for vanishing background Ramond-Ramnod fluxes the Wess-Zumino action of D6-branes does not contribute.
Let us next turn to the discussion of the $H^2 R$-sector and $(\nabla H)^2$-sector. The B-field  gives rise to real a scalar field $b = b(x)$ as
\beq
B = b \, \omega \,
\eeq
where as noted before $\omega$ is the unique harmonic $(1,1)$-form on $oY_3$. 
The results for the $H^2 R$-sector and $(\nabla H)^2$-sector \cite{Robbins_2014,Garousi_2017}  only hold for totally geodesic embeddings  i.e.\,for vanishing second fundamental form  $\Omega = 0$. However, a completion of the latter for generic embeddings would be required to perform the reduction relevant for our study. 
Nevertheless, one may  infer  the functional  form of  the $\alpha'{}^{5/2} g_s$-correction to the K\"ahler potential and coordinates from eq.\,\eqref{reduction-DBI_ODBI-D6}.
In the one-modulus case the K\"ahler potential is given by
\beq\label{Kpot1MD6}
K = \phi  - \, \text{log}\Big(  \cV + \kappa_1 \, \mathcal{Y} \, e^{\phi} \,  \cV^{\tfrac{1}{6}}\Big) \;\; ,
\eeq
and the complexified K\"ahler coordinates by
\ba\label{Kccord1MD6}
T &=  b +  i\, \Big(  \,  \cV^{\tfrac{1}{3}} +  \kappa_2  \, \mathcal{Y} \, e^{\phi}  \, \cV^{-\tfrac{1}{2}}  \Big) \;\; , \;\; \text{where}  \;\;\; \mathcal{Y} = \tfrac{1}{\sqrt{2 \pi}} \int_{D_m}  R_T |_{oY_3} \ast_3 1 \;\; ,
\ea
and with $\kappa_1,\kappa_2$ real parameters.\footnote{The scalar curvature  of $D_m$  generically is non-vanishing i.e.\,$R_T |_{oY_3} \neq 0$.
Moreover, note that  we have  factorized out  the volume one-modulus dependence in \eqref{AnsatzBGD5}. Thus  $R_T |_{oY_3}$ in the definition of $\mathcal{Y}$ is independent of the volume modulus.
} The classical form of  eq.\,\eqref{Kpot1MD6} and  eq.\,\eqref{Kccord1MD6}  are well known \cite{Grimm:2004ua,Grimm_2005}. Intriguingly, both the correction to the K\"ahler potential and the K\"ahler coordinates eq.\,\eqref{Kpot1MD6} and  eq.\,\eqref{Kccord1MD6} independently break the no-scale condition  as
\beq\label{no-scaleD6}
3 +  \frac{15 (\kappa_1 - 3 \kappa_2)}{ 4 \, \cV^{5/6}} \, e^{\phi} \,  \mathcal{Y}
\eeq
Let us close this section with a couple of  remarks.  Firstly note that although we expect $\mathcal{Y}$ to be a topological quantity such as the analog expressions on the four-cycle and two-cycle eq.\,\eqref{Z-def} and eq.\,\eqref{topXD5}, respectively, a proof of that proposition eludes us. Secondly, a comprehensive  study of the corrections of the b-scalar kinetic terms as well as the discussion of the corrected E.O.M.'s are required to fix the K\"ahler potential and coordinates. Only then one may infer if the no-scale condition \eqref{no-scaleD6} is broken at order $\alpha'{}^{5/2} g_s$. It would is of interest to analyze the impact of the $\alpha'$-correction in eq.\,\eqref{Kpot1MD6} on K\"ahler moduli stabilization in type IIA \cite{Palti_2008,McOrist_2012,escobar2018type}.

\subsection{Conclusions}
In this work we  have dimensionally reduced the next to leading order gravitational $\alpha'^2g_s$-corrections to the DBI actions of space-time filling D7-branes and a O7-plane on Calabi orientifold backgrounds with a single K\"ahler modulus. We found that the background solution of the dilaton, the warp-factor as well as the internal metric receive corrections. By studying the K\"ahler metric of the volume modulus  we found that either the K\"ahler potential or the K\"ahler coordinates or both  receive an  $\alpha'^2g_s$-correction which is of topological nature. Namely, carrying the  first Chern-form of the divisor wrapped by the D7's and O7. To draw definite conclusions one  is required to take into account the $F_5^2 R$ and $ ( \nabla F_5)^2$-terms to the DBI action which  however remain elusive. The latter could in principle be fixed by six-point open string disk and  projective plane amplitudes.

Finally  we established a connection of the results obtained  from the DBI actions in this work  to our previous F-theory results and found that the form of the  respective topological corrections is in agreement. The matching of the K\"ahler potential obtained in F-theory and the one from the DBI actions suggests that the no-scale structure is broken by the  $\alpha'^2g_s$-correction. Concludingly, we have obtained further evidence for the potential existence of the $\log \cV$-correction to the K\"ahler coordinates.  The search for an alternative interpretation as a loop effect to the D3-brane instanton action is  of great interest.

Moreover we have initiated the  study of  higher-derivative  corrections to the DBI action of space-time filling D5-branes and D6-branes - the latter in type IIA - on Calabi-Yau orientifold backgrounds with a single K\"ahler modulus and concluded that those potentially give rise to a novel $\alpha'^3 g_s$ and  $\alpha'{}^{5/2} g_s$-correction to the K\"ahler potential, respectively. However, a more extensive analysis is required to decide  upon their ultimate fate. 

To conclude,  let us emphasize that the ongoing quest to determine the leading order $\alpha'$-corrections to the  K\"ahler potential and coordinates of $4d \, \cN=1$ low energy theories in string theory is of great interest both for phenomenological as well as conceptual reasons, such as  their potential to generate the leading order perturbative scalar potential.


\newpage
\noindent{\bf Acknowledgements.}
Many thanks to Ralph Blumenhagen, Andreas Braun, Federico Bonetti, Andrew Macpherson, Michael Haack and Oliver Schlotterer for constructive  discussions and moreover Andreas Braun, Thomas Grimm, Michael Haack  and Kilian Mayer for their useful comments on the draft. In particular, let me take the opportunity to express my deep gratitude to Simeon Hellerman for  encouraging me to continue this research and moreover for our discussions  and for participating in an initial collaboration. Lastly, let me mention that I remain grateful for the support of my partner Sharon Law. The majority of this work was conducted while being affiliated with the Kavli IPMU at the University of Tokyo in 2019.
 This work was supported by the WPI program of Japan.

\noindent

\appendix

\section{\bf Conventions, definitions, and identities} \label{Conv_appendix}

In this work we denote the ten-dimensional space indices by capital Latin letters $M,N = 0,\ldots,10$ and
the external  ones by  $\mu,\nu = 0,1,2,3$ and internal ones $i,j = 1,\dots,6$. Furthermore,  the components in the tangent direction of the D7-branes  by $\alpha, \beta=0,...,7$ and the normal components by $a,b =1,2$. The metric signature of the eleven-dimensional space  is $(-,+,\dots,+)$.
Furthermore, the convention for the totally 
anti-symmetric tensor in Lorentzian space in an orthonormal frame is $\epsilon_{012...10} = \epsilon_{012}=+1$. 
We adopt the following conventions for the Christoffel symbols and Riemann tensor 
\ba\label{BinachiIds}
\G^R{}_{M N} & = \fr12 g^{RS} ( \pa_{M} g_{N S} + \pa_N g_{M S} - \pa_S g_{M N}  ) \, , &
R_{M N} & = R^R{}_{M R N} \, , \nonumber\\
R^{M}{}_{N R S} &= \pa_R \G^M{}_{S N}  - \pa_{S} \G^M{}_{R N} + \G^M{}_{R  T} \G^T{}_{S N} - \G^M{}_{ST} \G^T{}_{R N} \,, &
R & = R_{M N} g^{M N} \, , 
\ea
with equivalent definitions on the internal and external spaces. Written in components, the first and second  Bianchi identity are
\bea\label{Bainchiid}
{R^O}_{PMN} + {R^O}_{MNP}+{R^O}_{NPM} & = & 0 \nonumber\\
\nabla_L R^O{}_{PMN} + \nabla_M R^O{}_{PNL} + \nabla_N R^O{}_{PLM} & = & 0 \;\;\; .
\eea
Differential p-forms are expanded in a basis of differential one-forms as
\beq
\Lambda = \frac{1}{p!} \Lambda_{M_1\dots M_p} dx^{M_1}\wedge \dots \we dx^{M_p} \;\; .
\eeq
The wedge product between a $p$-form $\Lambda^{(p)}$ and a $q$-form $\Lambda^{(q)}$ is given by
\beq
(\Lambda^{(p)} \we \Lambda^{(q)})_{M_1 \dots M_{p+q}} = \frac{(p+q)!}{p!q!} \Lambda^{(p)}_{[M_1 \dots M_p} \Lambda^{(q)}_{M_1 \dots M_q]} \;\; .
\eeq
Furthermore, the exterior derivative on a $p$-form  $\Lambda$ results in
\beq
 ( d \Lambda)_{N M_1\dots M_p} = (p+1) \pa_{[N}\Lambda_{ M_1\dots M_p]} \;\;,
\eeq
while the Hodge star of   $p$-form  $\Lambda$ in $d$ real coordinates is given by
\beq
(\ast_d \Lambda)_{N_1 \dots N_{d-p}} = \frac{1}{p!} \Lambda^{M_1 \dots M_p}\epsilon_{M_1 \dots M_p N_1\dots N_{d-p}} \;\; .
\eeq
Moreover, 
\beq\label{idwestar}
 \Lambda^{(1)} \we \ast \Lambda^{(2)} = \frac{1}{p!}\Lambda^{(1)}_{M_1\dots M_p} \Lambda^{(2)}{}^{M_1\dots M_p} \ast_1 \;\; ,
\eeq
which holds  for two arbitrary $p$-forms $\Lambda^{(1)}$ and $\Lambda^{(2)}$.

Lastly, note that a Weyl rescaling of the four-dimensional metric
\beq\label{sec_Weylrescaling}
g'_{\mu \nu }\;\; = \;\; \Lambda g_{\mu \nu}  \;\; ,
\eeq
leads to a shift of the Ricci scalar  and the volume element as
\beq
R'= \tfrac{1}{\Lambda} R - \tfrac{3}{\Lambda^2} \nabla^\mu \nabla_\mu \Lambda +  \tfrac{3}{2\Lambda^3} \nabla^\mu \Lambda \nabla_\mu \Lambda \;\;\;\;\;\; ,\;\;\;\;  \ast'_4  \, 1 \;\;= \;\; \Lambda^2 \ast_4 \, 1  \;\; .
\eeq 
And furthermore that a rescaling of the  ten-dimensional Lorentzian metric such as $g'_{MN} = \Omega \, g_{MN} $ results in a shift of the Ricci scalar as 
\beq\label{10dWeyl}
R'= \tfrac{1}{\Omega} R - \tfrac{9}{\Omega^2} \nabla^M \nabla_M \Omega -  \tfrac{9}{\Omega^3} \nabla^M \Omega \nabla_M \Omega \;\;. 
\eeq
Let us next turn to discuss the Riemann tensor of the metric
\ba\label{Bgdeform}
ds^2 &= g^e_{\mu \nu}  dx ^\mu dx^\nu +  \cV ^{1/3}(x)   g_{ij} dy^i d y^j  \;\; , 
\ea
where $g_{ij} $ and $g^e_{\mu \nu} $ are the internal Calabi-Yau metric and external space metric, respectively. The components of the total space Riemann tensor are given by
\ba \label{deformationRiemann}
R_{ijkl} &= R(g)_{ijkl}   \cV ^{1/3} + \frac{1}{36 \cV^{4/3}} \, \big( g_{il} g_{jk} - g_{jl} g_{ik} \big)  \nabla_\mu \cV \,   \nabla^\mu \cV \;\; \\ \nonumber
R_{\mu i \nu j} &= \frac{5}{36 \cV^{5/3}} g_{ij} \nabla_\mu \cV \,   \nabla_\nu \cV  - \frac{1}{6 \cV^{2/3}} g_{ij} \, \nabla_\nu  \nabla_\mu \cV \;\; , \;\; 
R_{\mu \rho \nu \sigma} = R(g^e)_{\mu \rho \nu \sigma}
\ea
where $R(g)$ and $R(g^e)$ denote the Riemann tensor w.r.t. the Calabi-Yau metric and external metric, respectively.  Moreover,  $\nabla_\mu$ is the Levi-Civita connection of the external space metric. All other index combinations excpet for symmetries of the terms in \eqref{deformationRiemann} vanish.

\section{Immersions of D7-branes}\label{sec:Immersions}

\subsection{Geometry of sub-manifolds}
In this section we closely follow  \cite{Eisenhart1926,Kobayashi1969}. The embedding map of the D7-brane  into the ambient space is denoted by $Y^M$. A local frame of tangent vectors is given by $\partial_\alpha Y ^M$ and an orthogonal frame for the normal bundle by $\xi^M_a$ which obey per definition
\beq\label{tangent_normal_frame}
G_{M N}\partial_\alpha Y ^M \partial_\beta Y ^N = g_{\alpha \beta}  \;\;, \;\;\; G_{M N}  \xi^M_a \xi^N_b = \delta_{ab} \;\; , \text{and} \;\;\;\; G_{M N} \partial_\alpha Y ^M \xi^N_a = 0 \;\; ,
\eeq
where $g_{\alpha \beta}$ is the world-volume metric of the brane and $G_{MN} $ is the total space metric with its inverse given by
\beq\label{tangent_normal_frame_Inv}
G^{M N} = \partial_\alpha Y ^M \partial_\beta Y ^N  g^{\alpha \beta} + \xi^M_a \xi^N_b \delta^{ab} \;\;.
\eeq
 Note  that \ref{tangent_normal_frame}   and \ref{tangent_normal_frame_Inv}  imply that the metric is in the tangent and normal indices is of product form. 
 Note that  the tangent and normal frames are used to pull back indices form the total space e.g.
\beq
\hat R_{\alpha \beta \gamma \delta} = \hat R_{MNOP} \, \partial_\alpha Y ^M \partial_\beta  \, Y ^N \partial_\gamma \,  Y ^O \partial_\delta Y ^P \;\; \text{or} \;\;\;  \hat R_{ \alpha \beta a b}  =  \hat R_{MNOP}\, \partial_\alpha Y ^M \partial_\beta  \, Y ^N  \, \xi^{O}_a \, \xi^{P}_b \;\; .
\eeq
The second fundamental form is defined as
\beq
\Omega^{M}{}_{\alpha \beta} = \Omega^{M}{}_{\beta \alpha} = \partial_\alpha \partial_\beta Y^M - \Gamma_T{}^{\gamma}_{\alpha \beta} \partial{\gamma} Y^M + \Gamma^{M}_{NO} \partial_{\alpha} Y^N \partial_{\beta} Y^O \;\; .
\eeq
One may show that the tangent space projection  $\Omega^{\gamma}{}_{\alpha \beta}  = 0$ vanishes and thus the normals space projection
\beq
\Omega^{a}{}_{\alpha \beta}  \;\;,
\eeq
carries the entire information.
  For a minimal embedding one finds \cite{Eisenhart1926,Kobayashi1969,Takahashi1966} that
  \beq\label{Min_Omega}
   \Omega^{a} {}_{\alpha\beta} g^{\alpha \beta} = 0\;\;.
  \eeq
One may  also infer using the Ricci flatness of the Calabi-Yau orientifold  and in particular \eqref{tangent_normal_frame_Inv} that
\beq\label{CYO3-1}
\hat R_{\alpha c \beta}{}^c = - \hat  R_{\alpha \gamma \beta}{}^\gamma  \;\;,
\eeq
\beq\label{CYO3-2}
\hat R_{a c b }{}^c = -  \hat R_{a \gamma b}{}^\gamma  \;\;,
\eeq
\beq\label{CYO3-3}
\hat R_{a c \beta }{}^c = -  \hat R_{a \gamma \beta}{}^\gamma \;\;,
\eeq


\subsection{Minimal immersions  in Calabi-Yau manifolds}\label{minimal_Immersions}

In the following  we consider space-time filling BPS D7-branes which are wrapped holomorphic four-cycles of the internal Calabi-Yau space $Y_3$. Those four-cycles minimize the volume of the divisor $D_m$ in the Homology class i.e.\,the embedding map
 $\varphi : D_m \hookrightarrow Y_3 $ is a minimal immersion  \cite{Becker_1995}.  The same statement holds for the orientifolded Calabi-Yau $oY_3$. To avoid introducing yet other notation for indices we denote with $|_{oY_3}$ the restriction of object entirely to the Calabi-Yau space i.e.\,the index $\alpha,\beta = 1,\dots,4$  and $M=1,\dots,6$ on objects  dressed with $|_{oY_3}$.
One infers that  an isometric immersion is minimal if and only if the second fundamental form obeys 
 \beq 
 \Omega_{\alpha \beta }^{a} g^{\alpha \beta} |_{oY_3} = 0 \;\; , \;\; \;\; \Omega_{\alpha \beta }^{M} g^{\alpha \beta} |_{oY_3} = 0 \; ,  
 \eeq
 see e.g.\,\cite{Takahashi1966}.
 Also  note that for an isometric  immersion $\varphi$ into a Ricci flat space such as $Y_3$ of zero Ricci tensors and Ricci curvature one infers that
\beq\label{RicciFlatConnection}
\hat R_{a c \beta }{}^c |_{oY_3} = -  \hat R_{a \gamma \beta}{}^\gamma |_{oY_3}= 0 \;\; .
\eeq
Let us next discuss  the second fundamental form under the variation of the background metric w.r.t. the  internal volume  \eqref{Bgdeform}.  As we consider four-cycles in the Calabi-Yau geometry the only non-vanishing component is
\beq\label{RicciFlatConnection2}
\Omega_{\alpha \beta }^{a}=   \Omega_{\alpha \beta }^{a}  |_{oY_3}  \;\; ,
\eeq
where the l.h.s.\,describes the components of the eight-dimensional metric of the D7-brane  world-volume while the r.h.s.\,is our notation for the second fundamental form of the four-cycle embedding in the Calabi-Yau space. Note that in particular no derivatives terms of the Calabi-Yau volume are present. Where we have used that $\partial_\alpha Y ^M \sim \cV^{0}$ and $\xi^M_a \sim \cV^{-1/6}$.

Note that as in particular the minimal four-cycle  is a complex K\"ahler manifold one may use the properties of its K\"ahler metric. Such as that when expressed  in complex coordinates $z^\alpha, \bar z^{\bar \alpha}$,  with  $\alpha,\beta = 1,2$ and $\bar \alpha, \bar \beta = 1,2$  the metric is block-diagonal
\beq\label{complexKMetric}
g_{\alpha \bar \beta} = g_{\bar \beta \alpha }  \;,\;\; g_{\alpha \beta}  = 0 = g_{\bar\alpha \bar\beta}  \;\;,
\eeq
and that the components of the K\"ahler-form are given by
\beq
\tilde J_{\alpha \bar \beta} = \, -i \, g_{\alpha \bar \beta} \;\;.
\eeq
Moreover, the non-vanishing Riemann-tensor components are
\beq
R_T{}_{\alpha \bar \alpha \beta \bar \beta} |_{oY_3} \;\; ,
\eeq
where the Bianchi identity becomes  manifest i.e.\,the symmetric exchange of the holomorphic and anti-holomorphic indices, respectively.
 One defines the curvature two-form for Hermitian manifolds to be
  \begin{equation}\label{curvtwo}
 {\cR^\alpha}_\beta  =  {{R_T{}^\alpha}_\beta }_{ \gamma \bar \gamma}  |_{oY_3} dz^\gamma \wedge d\bar{z}^\bar{\gamma}\;\;,
  \end{equation}
 and 
 \bea \label{defR3}
 \Tr{\cR}\;\;&  =& {{R_T{}^\alpha}_\alpha }_{ \gamma \bar \gamma}  |_{oY_3} dz^\gamma \wedge d\bar{z}^\bar{\gamma} \;\; .
 \eea
 The first  Chern form can be expressed in terms of the curvature two-form as
\beq \label{DefFirstChern}
 c_1 \; =  \; \frac{i}{2 \pi} \Tr{ \mathcal{R}} \;\;.
\eeq



\subsection{Flux-background solution}\label{fluxdetails}
This appendix contains details of the flux-background. We refer the reader to section \ref{sec:SUSY-BG} for the 
comprehensive discussion.  We give the solution to
\beq
| F_3|^2_{\alpha \beta} , \; | H_3|^2_{\alpha \beta} \;\;  \sim \;\;  \mathcal{O}(\alpha)  \;\; \;\; \text{and} \;\;\;\; | F_3|^2_{a b} ,\; | H_3|^2_{ab} \;\;  \sim  \;\;\mathcal{O}(\alpha) \;\;,
\eeq
rather than $H_3$ and $F_3$. Note that many potential total derivative contributions are not fixed by the consistency with the Einstein equations and dilaton equation of motion. We use this freedom to to chose a particular representation of the flux components such that
\ba\label{fluxSolutionFinal}
 | H_3|^2_{ab} \; &=  | F_3|^2_{ab} = \mathcal{F}_{ab} \;\;, \\[0.2cm] \nonumber
  | H_3|^2_{\alpha \beta} &=   \mathcal{F}_{\alpha \beta}  + 6\, g^{\tiny (0)}_{\alpha \beta}  \, \nabla^{\tiny (0)}{}^a \nabla^{\tiny (0)}{}^b  \, \bar R_{ab} |_{oY_3} \;\;,  \\[0.2cm] \nonumber
   | F_3|^2_{\alpha \beta} &= e^{-2\phi_0} \big ( \mathcal{F}_{\alpha \beta}  - 6\, g^{\tiny (0)}_{\alpha \beta}  \, \nabla^{\tiny (0)}{}^a \nabla^{\tiny (0)}{}^b  \, \bar R_{ab} |_{oY_3} \big)\;\;, 
\ea
with
\ba\label{Fab}
\mathcal{F}_{ab}  \;\; =  \;\;  & \tfrac{6}{5} \, \nabla^{\tiny (0)}{}_a \nabla^{\tiny (0)}{}_b \, R_T  + 4  \, \nabla^{\tiny (0)}{}^c \nabla^{\tiny (0)}{}_{(a} \, \bar R_{b) c}  -  4  \, \nabla^{\tiny (0)}{}^c \nabla^{\tiny (0)}{}_c \, \bar R_{ab}
 \\[0.2cm] \nonumber
 & -  \gamma_3 \, \delta_{ab}\, \nabla^{\tiny (0)}{}^c \nabla^{\tiny (0)}{}_c \, R_T - 4  \, \delta_{ab}\, \nabla^{\tiny (0)}{}^c \nabla^{\tiny (0)}{}^d \, \bar R_{cd}  
  \\[0.2cm] \nonumber
  &+ (4 - \gamma_3) \, \delta_{ab}  \nabla^{\tiny (0)}{}^\alpha \nabla^{\tiny (0)}{}^\beta \, R_T {}_{\alpha \beta} \;\;,
\ea
and
\ba\label{Falbe}
 \mathcal{F}_{\alpha \beta} \;\;  =  \;\; &2\,  \nabla^{\tiny (0)}{}_a \nabla^{\tiny (0)}{}^a  \, R_T{}_{\alpha \beta} - \gamma_3 \, g^{\tiny (0)}_{\alpha \beta}  \,  \nabla^{\tiny (0)}{}_a \nabla^{\tiny (0)}{}^a  \, R_T  - 6\, \nabla^{\tiny (0)}{}^a \nabla^{\tiny (0)}{}^b  \, \bar R{}_{ab} \\[0.2cm] \nonumber
 &  + \tfrac{6}{5} \, \nabla^{\tiny (0)}{}_\alpha \nabla^{\tiny (0)}{}_\beta  \, R_T + 8\, \nabla^{\tiny (0)}{}^\gamma \nabla^{\tiny (0)}{}^\delta R_T{}_{\alpha \gamma\beta \delta} + 4\,\nabla^{\tiny (0)}{}^\gamma \nabla^{\tiny (0)}{}_{(\alpha }R_T{}_{\beta) \gamma} \\[0.2cm] \nonumber
&  -2 \,\nabla^{\tiny (0)}{}_\gamma \nabla^{\tiny (0)}{}^\gamma R_T{}_{\alpha \beta} - \gamma_3\, \, g^{\tiny (0)}_{\alpha \beta} \, \nabla^{\tiny (0)}{}^\gamma \nabla^{\tiny (0)}{}^\delta R_T{}_{\gamma \delta}  \;\;.
\ea
Let us emphasize that \eqref{fluxSolutionFinal} constitutes a particular choice where we have fixed the free undetermined parameters. Thus eq.'s  \eqref{Fab} and \eqref{Falbe} remain to  depend only on the parameter $\gamma_3$.

\nocite{*}
\bibliographystyle{utcaps}
\newpage
\bibliography{references}
\end{document}